\documentclass[structabstract]{aa}
\usepackage{graphicx}
\usepackage{epstopdf}
\usepackage{natbib}
\usepackage{color}
\bibpunct{(}{)}{;}{a}{}{,}   

\newcommand{\vect}[1]{\mathbf{#1}}   
\newcommand{\beq}{\begin{equation}}
\newcommand{\eeq}{\end{equation}}
\newcommand{\half}{\frac{1}{2}}

\usepackage{txfonts}
\begin{document}
   \title{Simulation of the growth of the 3D Rayleigh-Taylor instability\\in supernova remnants using an expanding reference frame}
   \titlerunning{Growth of the 3D Rayleigh-Taylor instability in Supernova Remnants using expanding reference frames}
\authorrunning{Fraschetti F.}

\author{Federico Fraschetti$^{1,2,3}$, Romain Teyssier$^{1,4}$, Jean Ballet$^{1}$, \and Anne Decourchelle$^{1}$} 

\institute{$^{1}$Laboratoire AIM, CEA/DSM - CNRS - Universit\'e Paris Diderot,
IRFU/SAp, F-91191 Gif sur Yvette, France\\$^{2}$LUTh, Observatoire de Paris, CNRS-UMR8102 and Universit\'e Paris VII,
5 Place Jules Janssen, F-92195 Meudon C\'edex, France.\\$^{3}$Lunar and Planetary Lab \& Dept. of Physics, University of Arizona, Tucson, AZ, 85721, USA.\\$^{4}$Institute for Theoretical Physics, University of Zurich, CH-8057 Zurich, Switzerland .\\
              \email{federico.fraschetti@cea.fr}
              }

   \date{Received .........; accepted .....}


  \abstract
   {The Rayleigh-Taylor instabilities that are generated by the deceleration of a supernova remnant during the ejecta-dominated phase are known to produce finger-like structures in the matter distribution that modify the geometry of the remnant. The morphology of supernova remnants is also expected to be modified when efficient particle acceleration occurs at their shocks.}
   {The impact of the Rayleigh-Taylor instabilities from the ejecta-dominated to the Sedov-Taylor phase is investigated over one octant of the supernova remnant. We also study the effect of efficient particle acceleration at the forward shock on the growth of the Rayleigh-Taylor instabilities.}
   {We modified the Adaptive Mesh Refinement code RAMSES to study with hydrodynamic numerical simulations the evolution of supernova remnants in the framework of an expanding reference frame. The adiabatic index of a relativistic gas between the forward shock and the contact discontinuity mimics the presence of accelerated particles.}
   {The great advantage of the super-comoving coordinate system adopted here is that it minimizes numerical diffusion at the contact discontinuity, since it is stationary with respect to the grid. We propose an accurate expression for the growth of the Rayleigh-Taylor structures that smoothly connects the early growth to the asymptotic self-similar behaviour.}
   {The development of the Rayleigh-Taylor structures is affected, although not drastically, if the blast wave is dominated by cosmic rays. The amount of ejecta that reaches the shocked interstellar medium is smaller in this case. If acceleration were to occur at both shocks,  the extent of the Rayleigh-Taylor structures would be similar but the reverse shock would be strongly perturbed.}
 
   \keywords{ISM: supernova remnants - Physical data and processes: Acceleration of particles, Hydrodynamics}

   \maketitle
%

\section{Introduction}\label{int}
In young supernova remnants (hereafter SNR), the dense shell of material ejected by the explosion and decelerating in a rare\-fied interstellar medium (hereafter ISM) is expected to be subject to hydrodynamic instabilities \citep{g73,g75,s78} of Rayleigh-Taylor (hereafter RT) type.
These instabilities modify the morphology of the SNR causing a departure of the ejecta from spherical symmetry. They manifest themselves as finger-like structures of material protruding from the contact discontinuity between the two media into the ISM heated by the forward shock, as shown by numerical simu\-lations \citep[e.g.][]{cbe92,dc98,d00,wc01}. During this process, the shocked ejecta and the shocked ISM remain two distinct fluids. The X-ray observations of Tycho's SNR \citep{w05} exhibit structures that are consistent with these effects. Deviations from spherical symmetry may also be caused by initial asymmetries in the explosion of the progenitor or local inhomogeneities in the circumstellar medium. In Cassiopeia A, the spatial inversion observed by the Chandra observatory of the iron and silicon layers \citep{hrbs00} provides a strong indication of an asymmetric explosion. An even more radical example of deviation from spherical symmetry is SN 1993J, a stellar wind case, where the optical and radio observations can be reconciled by assuming a strong asphericity for the pre-existing progenitor activity \citep{bbr01}. 

Two-dimensional hydrodynamic simulations of SNRs can take into account the RT instability. Three-dimensional simulations have shown an enhancement of small-scale structures and more severe deformation of the reverse shock surface \citep{jn96}; the perturbation seems to grow faster by $30\%$ than in the two-dimensional case \citep{k00}. We chose to pursue three-dimensional hydrodynamic numerical simulations, focusing on the deviations from spherical symmetry of the SNR in the ejecta-dominated phase induced by the RT instabilities.

Several distinct physical processes occur in young SNRs, for instance, the dispersion of synthesized materials in the circumstellar medium and the propagation of collisionless shocks. Galactic SNRs are also considered to be a strong candidate source of galactic cosmic rays up to the knee, i.e. $E \sim 3\times10^{15}$ eV \citep{lc83}, since the rate and energy budget can account for the galactic energy density of cosmic rays; however, a direct identification of SNRs as sources of cosmic-ray nuclei is still lacking.

In the present paper, we aim to investigate with hydrodynamic equations the growth of Rayleigh-Taylor instabilities in the context of efficient particle acceleration. We adapt to SNR physics the hydrodynamic version of the AMR (Adaptive Mesh Refinement) code RAMSES \citep{t02}, designed originally to study the large-scale structure formation in the universe with high spatial resolution. The first application, considered here, is to study the effect of the growth of RT instabilities on the profile of the hydrodynamic variables by considering the evolution of a full octant of the SNR, i.e. a larger angular region than previously considered \citep{be01}. We do not introduce any seed perturbation into the 3D radial velocity field, so that any departure from spherical symmetry is naturally produced by the numerical fluctuations; this is supported by the non-linear phase of the instability being insensitive to initial perturbations \citep{cbe92}. 

\citet{be01} described the efficient acceleration of particles by changing the effective adiabatic index throughout. A higher compression ratio at the shocks was found to only slightly affect the growth of RT instabilities. However, the suggestion that the reverse shock can efficiently accelerate particles continues to be debated \citep{edb05}. In this paper, we study the impact of cosmic-ray acceleration on the instabilities using a simplified description that assumes that the adiabatic index is 4/3 in the shocked ISM region, but remains 5/3 inside the contact discontinuity (surface separating ejecta and ISM). This simulates the case of a cosmic ray-dominated blast wave and gas-dominated reverse shock.

To follow the expansion of SNRs, we use synergically two numerical approaches: AMR and the Moving Computational Grid. As a result, the large-scale turbulence in SNR can be simu\-lated, allowing for an accurate description of the instabilities.

The hydrodynamic equations can be formulated with respect to two distinct classes of coordinate systems: Eulerian, i.e., fixed space-coordinates in time, whose major concern is the production of numerical diffusion due to the advective terms; and Lagrangian, i.e., coordinates comoving with the bulk fluid, which is free in principle from numerical diffusion, but possible grid distortions require a rezoning which produces new numerical diffusion. Therefore, Eulerian coordinate systems are generally selected for multidimensional flow simulations.  
The numerical approach of Moving Computational Grid consists of using a computational grid comoving, or quasi-comoving, with the hydrodynamic flow to minimize the local fluid velocity. The idea of a computational grid adapted to follow the global motion of the fluid was first proposed in cosmological numerical simu\-lations by \citet{g95}. However, further analysis \citep{gb96} highlighted problems in the coupling of the hydrodynamics solver with the gravity solver. Strong mesh deformations were also found in the gravitational clustering. In the approach of Moving Computational Grid, the mesh moves continuously and in addition to a full transformation of coordinates (position and time), a transformation of hydrodynamic variables (density, velocity and pressure) is performed. In contrast, in a purely Lagrangian approach the space-coordinates are comoving with the bulk flow, leaving the hydrodynamic quantities unchanged.

The main disadvantage of simulating a SNR in a Eulerian fixed computational grid is that in the bulk flow of the SNR the total energy is roughly equal to the kinetic energy. Therefore, the thermal energy, computed as the difference between total and kinetic energies, can be very small leading possibly to local negative thermodynamic pressure. An algorithm is required to control the sign of pressure.

We apply a combination of the AMR with the Moving Computational Grid from the {\it young} phase, starting soon after the self-similar profile of \citet{c82,c83} has been established. The modification introduced enabled us to follow the evolution of one eighth of the volume of a young SNR, a large volu\-me with respect to previous 3D simulations \citep[e.g.][]{jn96}, and for a long interval of time, namely until the transition to the Sedov-Taylor phase. We do not provide all the details of the RAMSES code, which can be found in \citet{t02}, but a description of the modifications introduced here is outlined. 

The plan of the paper is the following. In Sect.~\ref{inicond}, we discuss the initial conditions adopted for a young SNR, namely the mapping of the uni-dimensional solution of the hydrodynamic equation over a cartesian grid. In Sect.~\ref{eqs}, we present the  hydrodynamic equations for the SNR flow, written in both the laboratory frame and the reference frame which is  comoving with the contact discontinuity. In Sect.~\ref{num}, we sketch the main steps of the implementation of the Godunov method in RAMSES, we discuss the modifications and the new variables introduced: the active variable $\alpha$ to change locally the equation of state; the passive scalar $f$, which traces the surface of the contact discontinuity; and the ionization age $\tau$. In Sect.~\ref{snr}, we present the results of SNR simulations: the growth of the RT structures and the elongation in time; the influence of a cosmic-ray-dominated blast wave on the RT instabilities. In Sect.~\ref{disc}, we discuss the main findings and present additional numerical tests. In Sect.~\ref{conc}, we conclude and present forthcoming applications.

\section{Initial conditions for self-similar expansion of young SNRs}\label{inicond}

Supernovae observations have confirmed \citep[for the case of SN 1987A, see][]{a88} that at early times, typically a few hours after the supernova explosion, the matter density of the outer ejecta reaches a power-law profile as a function of radius given by $\rho = \rho_{0} (r/r_0)^{-n}$. A kinematical study of the pre-Sedov-Taylor SNR of Kepler \citep{v08} confirmed the power-law density profile. The difficulties in attaining a measurement of $n$ makes the comparison of theory with observations difficult. The value of $n$ depends on the properties of the progenitor star and the acceleration of the shock during the explosion. The value $n=7$, or alternatively an exponential density profile for ejecta \citep{dc98}, is commonly used to describe Type Ia supernovae, which are believed to result from the explosion of a C/O white dwarf accreted in a binary system \citep{hf60}. The value $n=9$ is believed to describe Type II supernovae, which result from the core-collapse of a massive star. If the density of the ambient medium has a power-law behaviour, namely $\rho = \rho_{\rm ISM} (r/r_0)^{-s}$, self-similar solutions can be found for the density, velocity, and pressure of the interaction region in young SNRs \citep{c82}; from dimensional analy\-sis, the time evolution of the radius of the contact discontinuity has the form \citep{c82} $r(t) \propto t^\lambda$, where $\lambda = (n-3)/(n-s)$. 

During the expansion, a colder and denser fluid, the shocked ejecta, slows down in a hotter and less dense fluid, the shocked ISM. Seen in the reference frame of the contact discontinuity, this deceleration is equivalent to a gravity field pushing the shocked ejecta toward the shocked ISM. This instability is at the basis of the familiar RT structures \citep{cbe92}. The structure of the interaction region between the ballistically expanding ejecta and a uniform ISM in young SNR was first explored numerically by \citet{g73,g75}, who described how the formation of optical filaments is brought back to the RT instability and provided a qualitative explanation of the amplification of the magnetic field in the interaction region. Given spherical initial conditions, we investigate the growth of RT instabilities in different situations.

In this paper, a steep power-law density ejecta is assumed to expand into a uniform ISM ($s=0$).
In the innermost part, the common cutoff in radius $r_{\mathrm c}$ is adopted to introduce the density plateau for the ejecta core
\begin{equation}
\rho_{\rm ej}(t, r) = \left\{
  \begin{array}{cc}
    \rho_{\mathrm c} (t) (r/r_{\mathrm c})^{-n} & \mathrm{if~} r > r_{\mathrm c} \, \\
    \rho_{\mathrm c} (t) & \mathrm{if~} r < r_{\mathrm c}\, ,
    \label{rho_c}
   \end{array}
\right.
\end{equation}
where $ \rho_{\mathrm c} (t) = g^n t^{n-3}  r_{\mathrm c} ^{-n} \propto t^{-3}$, and $g^n$ is the normalization constant in \citet{c82}. As expected, $\rho_{\mathrm c} (t)$ dilutes from the initial high values in the inner core down to a negligible value during the Sedov-Taylor phase. 

At a given time $t_{\mathrm 0}$, the physical parameters that fix unequivo\-cally the profiles of $\rho$, $\vect{u}$, and $P$ are the total kinetic energy $E_{\mathrm 0}$ of the explosion, the total ejecta mass $M_{\mathrm {ej}}$, the circumstellar medium density $\rho_{\rm ISM}$ and the indices of the power-law density profile of ejecta and ISM, namely $(n,s)$. The outer radius of the inner plateau is given by  
\begin{equation}
r_{\rm c} = t \sqrt{\frac{10}{3}\frac{n-5}{n-3}\frac{E_0}{M_{\rm ej}}}. 
\end{equation}
A SNR can be defined to be young when the mass of interstellar matter $M_{\rm ISM}$ swept up is a small fraction of the mass of ejecta $M_{\rm ej}$ given by $M_{\rm ISM} < M_{\rm ej}$. A more quantitative criterion is that the young SNR phase finishes when the reverse shock radius $r_{\rm RS}$ has reached the radius $r_{\rm c}$ of the core of the remnant, 
 
The self-similar profiles of mass density $\rho$, fluid velocity $\vect{u}$ and pressure $P$ for a young SNR \citep{c82,c83} are used as initial conditions (see Fig.~\ref{ini}). By assuming spherical symmetry, the self-similar solution is mapped in a 3D computational grid. At the initial time $t_{0} = 10$ yr, the self-similar profiles are assumed to be well-established. It is reasonable to assume that such a spherical symmetry has not yet been significantly modi\-fied by the convective motions grown between the explosion of the supernova and the time $t_{0}$. Given the value of $t_{0}$, we can fix $r_{\rm c} (t_0)$ and therefore the density profile is determined down to $r=0$, where the origin of the explosion is located. We assume the reasonable values of SNR parameters $E_0 = 1.6 \times 10^{51}$ erg and $M_{\rm ej} = 5.0 \times M_{\odot}$, where $M_{\odot} = 1.989\times 10^{33}$ g is the solar mass and the circumstellar medium density $\rho_{\rm ISM} = 0.42$ amu$\times$cm$^{-3}$. The only information concerning the progenitor of the SNR is therefore contained in the index $n$ and the mass $M_{\rm ej}$. For the previous set of parameters, we find the initial radii of the contact discontinuity, forward shock and reverse shock to be, respectively, $r_{\rm CD} = 0.348$ pc, $r_{\rm FS} = 0.412$ pc, and $r_{\rm RS} = 0.326$ pc. 

\section{Equation of hydrodynamics in accelerated frame}
\label{eqs}
\subsection{Euler equations in laboratory frame}\label{eqs1}

In an inertial cartesian (laboratory) reference frame $\mathbf{R} = \{t ,\mathbf{r} \}$, the non-relativistic Euler equations of fluid dynamics, in the absence of viscosity and dissipative processes, are written in the quasi-conservative form
\begin{eqnarray}
\frac{\partial}{\partial t}\pmatrix{\rho \cr \rho u_{i} \cr E \cr \alpha}+ \nabla_{j} \pmatrix{\rho u_{j} \cr \rho u_{i} u_{j} + \delta_{ij}P \cr u_{j}(E + P) \cr \alpha u_{j}} = \pmatrix{0 \cr 0 \cr 0 \cr \alpha \nabla_{j} u_{j}} \, ,
\label{Euler}
\end{eqnarray}
where $\rho$ is the mass density, $u_{i}$ is the $i$-th component of fluid velocity, $P$ is the pressure, $E = \epsilon  + \frac{1}{2}\rho u^2$ is the total energy density, and $\epsilon$ is the internal energy density. The solution of these equations requires the use of an equation of state to find the quantity $P/\rho$ given the quantity $\epsilon/\rho$ \citep{s57}. We use the common short-cut for the adiabatic case, consisting of defining the parameter $\gamma$ by means of the equation $P  =  (\gamma - 1) \epsilon$. In the non-relativistic case, $\gamma$ can be identified as the ratio of the specific heats of the gas ($\gamma = 5/3$). If the pressure is dominated by the relativistic particles, $\gamma = 4/3$. The active scalar $\alpha = (\gamma - 1)^{-1}$ is introduced to treat in a purely hydrodynamic way the effect of efficient particle acceleration at the forward shock (see Sect.~\ref{nonthermal}).
 
Equation~\ref{Euler} is written in conservative form, except the so-called ``quasi-conservative'' form of equation for $\alpha$. If the conservative variable $\rho \alpha$ is used, higher order schemes of shock capturing applied to multicomponent flows can produce oscillations at the interface of different fluids \citep{s98,jc06}.

For a shock propagating into an ideal gas, the compression factor at the shock can be expressed as $\rho_2/\rho_1= (\gamma +1)/(\gamma - 1 +2/M_1 ^2)$ \citep{zr02}, where $\rho_2$ ($\rho_1$) is the downstream (upstream) density of the shock and $M_1 =  [\rho_1u_1^2/\gamma_1 P_1]^{(1/2)}$ is the Mach number in the unshocked medium. In the limit of a very strong shock, as in young SNRs, $M_1 \rightarrow \infty$ and $\rho_2/\rho_1 \rightarrow (\gamma +1)/(\gamma - 1)$, depending only on the thermodynamic properties of the medium.

The passive scalar $f$, representing the mass fraction of ejecta, is initially defined as
\begin{equation}
f(r) = \left\{
  \begin{array}{cc}
    1 & \mathrm{if~} r < r_{\mathrm{CD}} \, ,\\
    0 & \mathrm{if~} r > r_{\mathrm{CD}}\, ,
    \label{fp}
   \end{array}
\right.
\end{equation}
where $r_{\rm CD}$ is the radius of the contact discontinuity. The gradient of the function $f$ traces the surface of contact discontinuity,  namely the finger-like structures arising because of the RT instabilities, and $f$ is propagated by means of
\begin{equation}
\frac{\partial}{\partial t} (\rho f) + \nabla_{j} (\rho f u_{j}) = { 0} \, .
\label{eq_fp}
\end{equation}

The X-ray spectra of SNRs usually exhibit significant emission lines of heavy elements, such as Fe in Kepler \citep{spabrr89}. Therefore, the computation of the X-ray spectrum of SNRs requires an evaluation of the ionization produced by electrons after the passage of the reverse shock. In this paper, we do not compute the ionization state by coupling the hydrodynamics equations with the equations for the evolution of ionization and recombination \citep[for a 1D example of this see][]{r94}. We follow the ionization age, which allows us to compute a posteriori the ionization structure assuming a constant temperature in the fluid element. 

An additional passive scalar representing the ionization age, $\tau$, was introduced into the code, defined as
\begin{equation}
\tau(t,\vect{r}) = \int_{t_{\rm sh}} ^t n_{\mathrm e} (t',\vect{r}) dt' \, ,
\label{taudef} 
\end{equation}
where $t_{\rm sh}$ is the time since crossing the forward or reverse shock at point $\vect{r}$ and $n_{\mathrm e}(t,\vect{r})$ is the number density of post-shock electrons in the heated plasma. As a passive scalar, $\tau$ does not cause any variations in the dynamics of the SNR. The ionization age $\tau$ satisfies the equation
\begin{equation}
\frac{\partial}{\partial t}(\rho\tau)+ \nabla_{j} (\rho \tau u_{j}) = {\rho n_\mathrm{e} \vartheta(P(\vect{r}) - B P_{\rm ISM})} \, ,
\label{eq_tau}
\end{equation}
where $\vartheta(x)$ is the Heaviside function, $P_{\rm ISM}$ is the pressure in the unshocked ISM, and we can assume for the pressure $P(\vect{r})$ in the free expansion region, i.e., $r < r_{\rm RS}$, that $P(\vect{r}) < P_{\rm ISM}$, therefore Eq.~\ref{eq_tau} is homogeneous outside the interaction region. The constant $B$ is the threshold ratio for the detection algorithm of the interaction region. Since the pressure $P_{\rm int}$ in the interaction region satisfies $P_{\rm int} \gg P_{\rm ISM}$, we can choose $B=10^{2}$. No failures in the detection algorithm have been found for lower values of $B$ down to $B=10$.

\subsection{Euler equations in comoving reference frame}

Now we consider a non-inertial reference frame $\mathbf{\tilde R} = \{\tilde t,\vect{\tilde r} \}$. We deal with the simplified case of a purely radially contracting accelerated motion, which was introduced in \citet{ms98} as supercomoving coordinate system; the more general case in the presence of rotation is treated in \citet{pk07}. The frame $\mathbf{\tilde R}$ is defined by the transformation $\mathbf{\Lambda}: \mathbf{R} \mapsto \mathbf{\tilde R}$
\begin{eqnarray}
\vect{\tilde r}  & = &  \displaystyle \vect{r}/a \, ,
\label{Euler1tilde} \\
d{\tilde t}   & = & \displaystyle \frac{dt}{a^{\beta + 1}} \, ,
\label{Euler2tilde} \\
\tilde \rho(\vect{\tilde r},\tilde t) & = & a^{\phi}\rho\,  ,
\label{Euler3tilde}
\end{eqnarray}
where $\phi$ and $\beta$ are two scaling parameters. The scale factor $a(t)$ must be a non-vanishing twice-differentiable function of time only. Once the transformation law of the three fundamental quantities, namely length, time and mass, is fixed, the tranformation law of all the others can be found. From Eqs.~\ref{Euler1tilde} and \ref{Euler2tilde}, we obtain the transformation law of the velocity vector
\beq
\vect{\tilde u} = a^{\beta} \left(\vect{u} - \frac{\dot a}{a}\vect{r} \right) \, ,
\label{newvel}
\eeq
where $\dot a = da/dt$. The velocity $\vect{\tilde u}$ is decomposed into the veloci\-ty in the inertial frame $\mathbf{R}$ and the relative velocity of the bulk motion. To establish a computational grid comoving with the hydrodynamic flow, we choose the expansion law 
\beq
a(t) = a_0 (t/t_0)^\lambda \, ,
\label{at}
\eeq
where we consider the simple case of $\lambda$ being independent of time. In the cosmological framework, the transformation to supercomoving coordinates maps the uniform and isotropic Hubble flow in a matter-dominated universe to a uniform matter distribution at rest with constant thermodynamic properties, in the adiabatic case, namely $\gamma = 5/3$. In that case, the aim is to focus on deviations from the Hubble flow that drive the structure formation in the universe. In the case of self-similar expansion of SNR, the transformation of \citet{ms98} has the effect of converting the mesh from Eulerian to quasi-Lagrangian, which is exactly comoving with the contact discontinuity while the self-similar expansion applies, but does not experience grid distortions as in the case of purely Lagrangian mesh. 

From Eqs.~\ref{Euler1tilde} and \ref{Euler2tilde}, we also have
\begin{eqnarray}
\tilde P(\vect{\tilde r},\tilde t) & = & a^{\phi +2\beta}P  \\
\tilde \epsilon(\vect{\tilde r},\tilde t) & = & a^{\phi +2\beta} \epsilon .
\label{pressure}
\end{eqnarray}
In 3D, the choice $\phi = 3$, for any value of $\beta$, preserves the invariance of the mass conservation equation. However, as shown explicitly in \citet{pk07}, only the choice
\begin{equation}
\phi = 3 ,\quad \beta = 1
\label{values}
\end{equation}
and $\dot a = const$ for a perfect monoatomic gas with $\gamma = 5/3$ preserves the invariance of the Euler equation (Eq.~\ref{Euler}). In this paper, we report results obtained by fixing $\phi$ and $\beta$ in Eq.~\ref{values} and $a(t) = a_0 (t/t_0)^\lambda$, fixing the value of $\lambda$ according to the physical parameters of the specific problem. Hereafter, the choice $a_0 = 1$ is applied without reducing the generality of the treatment. In the reference frame $\mathbf{\tilde R }$, the Euler equations (Eq.~\ref{Euler}) become
\begin{eqnarray}
\frac{\partial}{\partial {\tilde t}}\pmatrix{{\tilde \rho} \cr {\tilde \rho} {\tilde u}_i \cr \tilde E \cr \alpha } + \tilde\nabla_{j} \pmatrix{{\tilde \rho} {\tilde u}_{j} \cr {\tilde \rho} {\tilde u}_i {\tilde u}_{j} + \delta_{ij} {\tilde P} \cr {\tilde u}_{j}(\tilde E + \tilde P) \cr \alpha {\tilde u}_{j} } =
\pmatrix{0 \cr  -{\tilde \rho} {\tilde r_i} {\mathcal G} \cr - {\tilde \rho}({\tilde r_j} {\tilde u_j}){\mathcal G}  + {\tilde H} {\tilde \epsilon} (5 -  3\gamma) \cr \alpha \tilde\nabla_{j} {\tilde u}_{j} } \nonumber \, , \\
\label{Eulertilde}
\end{eqnarray}
where $\tilde E = \tilde\epsilon  + \frac{1}{2} \tilde \rho {\tilde u}^2$ , ${\tilde H} = \frac{1}{a}\frac{da}{d\tilde t}$ and
\beq
{\mathcal G} = \frac{1}{a} \frac{d^2 a}{d{\tilde t}^2} -  \frac{2}{a^2} \left( \frac{da}{d \tilde t}\right)^2 = a^3 \frac{d^2 a}{d {t^2}}.
\label{G}
\eeq
Here, for a generic function $g(t, \vect{r})$ we have
\begin{eqnarray}
\left({\frac{\partial g}{\partial t}}\right)_\vect{r} &=& {1\over a^2}\left[\left({\frac{\partial g}{\partial\tilde t}}\right)_\vect{r} - {\tilde H}\vect{r} \cdot(\tilde \nabla g)_{\tilde t}\right]\,,\\
(\nabla g)_t &=& {1\over a}(\nabla g)_{\tilde t}\,.
\end{eqnarray}
In most physical cases, the bulk motion is accelerated; therefore in the momentum equation written in the non-inertial reference frame a non-inertial force term will appear on the right-hand side. In the right-hand side of Eq.~\ref{Eulertilde}, both the second line and the first term of the third line represent the non-inertial force; the second term of the third line, treated as a source term since it acts as the gravitational potential term in the cosmological framework, may create an energy sink, losing the numerical advantage of the conservative form (see Sect.~\ref{num} for details). 

Since it is an advection equation, Eq.~\ref{eq_fp} is unchanged by the choice of Eq.~\ref{values}
\begin{equation}
\frac{\partial}{\partial {\tilde t}}{(\tilde \rho} f ) + \tilde\nabla_{j} { (\tilde \rho} f {\tilde u}_{j} ) = 0 . \,
\label{Eulertilde_fp}
\end{equation}
Equation~\ref{eq_tau} changes as follows
\begin{equation}
\frac{\partial}{\partial {\tilde t}}{(\tilde\rho \tau) } + \tilde\nabla_{j} { (\tilde \rho} \tau {\tilde u}_{j} ) =
{ \tilde\rho {\tilde n}_e a^{-1}\vartheta(\tilde P(\vect{\tilde r}) - B \tilde P_{\rm ISM})} .
\label{Eulertilde_tau}
\end{equation}
In this paper, we consider the early phase of SNR expansion and the transition to the Sedov-Taylor phase, thus the radiative cooling is negligible with respect to the total energy content and the approximation of adiabatic expansion of a perfect fluid is justified. We assume that the fluid mainly consists of fully ionized hydrogen, therefore $\gamma = 5/3$. We notice that for $\gamma = 5/3$ the last term in the third relation of Eq.~\ref{Eulertilde} disappears. If $\gamma \neq 5/3$, as in Sect.~\ref{nonthermal}, this term can be regarded as a cooling or heating term, representing a microscopic form in which the internal energy of a non-monoatomic perfect gas can be deposited in the fluid as vibro-rotational degrees of freedom. 

The physical time $t$ is related to the time $\tilde t$ for $\phi=3$ and $\beta = 1$ by 
\begin{equation}
\tilde t = \frac{t_0}{a_0 ^2} \frac{1}{2\lambda - 1}\left[ 1 - \left(\frac{t_0}{t} \right)^{2\lambda - 1} \right] \quad.\
\label{ttilde}
\end{equation}
In the cases considered, $\lambda \ge 4/7$ ($n \ge 7$ and $s \ge 0$), thus the factor $(2\lambda - 1)^{-1} > 0$ for every $\lambda$. Therefore, the evolution in $\tilde t$ ranges from $\tilde t = 0$, for $t = t_0$, to $\tilde t \to (t_0/a_0 ^2) (2\lambda - 1)^{-1}$ for $t \to \infty$. This squeezing of the total time interval depends on the parameters of the progenitor of the SNR.
We also note that, because of Eq.~\ref{Euler2tilde}, since $\beta > 0$, the time in $\mathbf{\tilde R}$ is compressed relative to the physical time in $\mathbf{R}$ by a factor $a^{\beta + 1}$, reducing the time step of the numerical integration.

\begin{center}
\begin{figure}
\includegraphics[width=9.4cm]{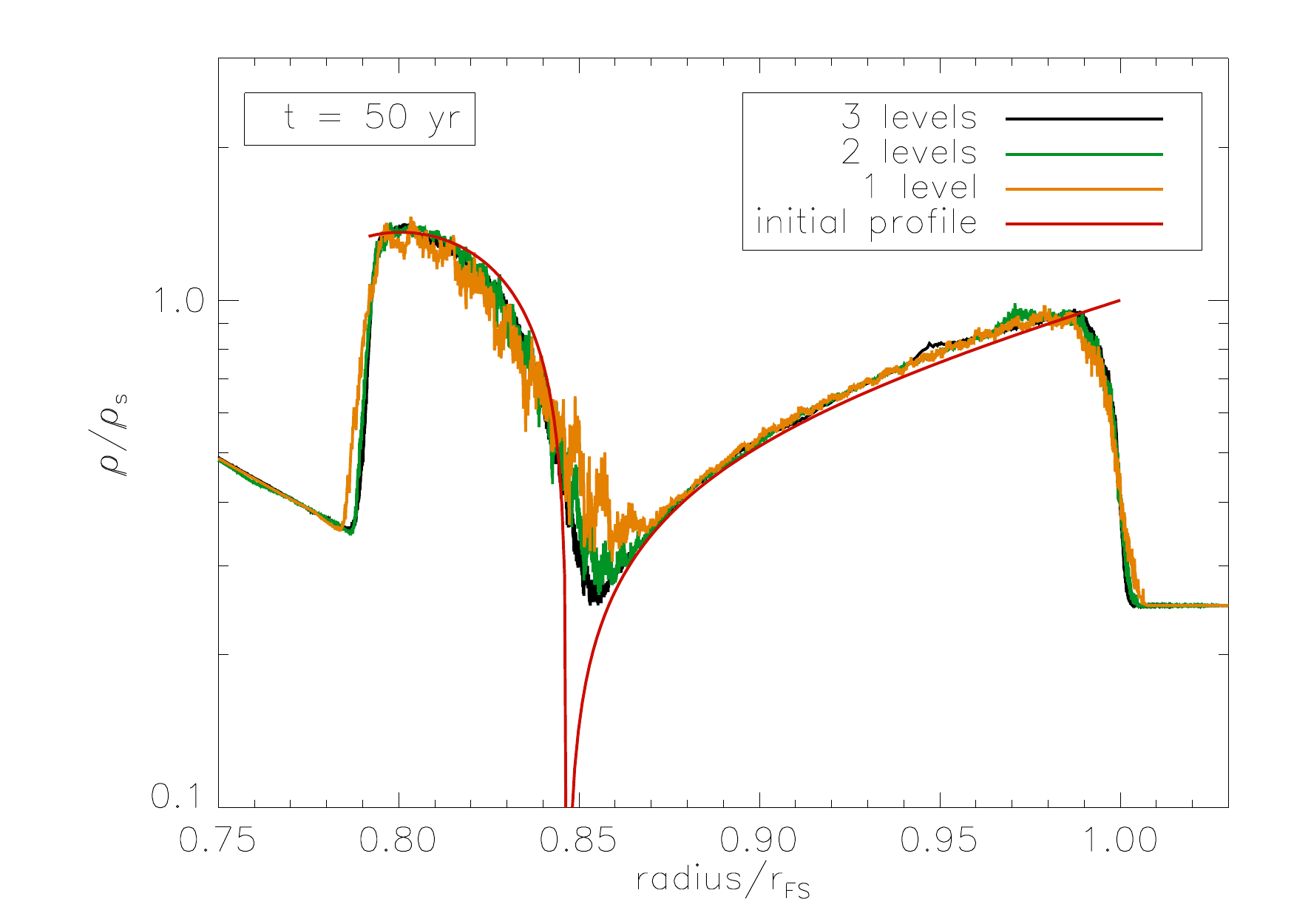}
\caption{Angle-averaged radial density profile at $t=50$ yr for $\lambda = 0.57$, or $(n,s) = (7,0)$, and $\gamma = 5/3$ with 1, 2, and 3 levels of refinement ($128^3$ in yellow, $(128 - 256)^3$ in green, $(128 -512)^3$ in black). For comparison, the self-similar profile is shown in red. The density profiles are normalized to their values at the forward shock. From this diagram, it can be seen that the localization of the forward shock and its compression ratio are not strongly dependent on the maximal refinement level.  \label{ini}}
\end{figure}
\end{center}

\section{Numerical code}\label{num}

\subsection{Set-up}\label{setup}

The simulations were performed with the hydrodyna\-mics version of the code RAMSES \citep{t02}. RAMSES  was origi\-nally developed to study, with the AMR technique, structure formation in the universe at high spatial resolution. The solver is based on a second order Godunov scheme. We modified the constitutive equations of RAMSES as shown in Eqs.~\ref{Eulertilde} and \ref{G} to solve the hydrodynamics flow in an accelerated reference frame expanding according to Eqs.~\ref{Euler1tilde}, \ref{Euler2tilde}, and \ref{Euler3tilde}. The modifications concern the hydrodynamic solver routines and the external gravity routine, because the change of reference frame introduces a non-inertial force in the Euler equations, which can be treated as a source term in the momentum and energy equation. We also introduced an additional variable $\alpha$, treated as a passive scalar in Eqs.~\ref{Euler} and~\ref{Eulertilde}, to change locally the equation of state and two more passive scalars (see Sect.~\ref{eqs1} for details).

The simulations were performed over three levels of refinement, of value $l=7-9$, corresponding to a cartesian grid with a number of cells $128^3$, $256^3$, and $512^3$, respectively. The effective maximal resolution attained in the interaction region at the set-up of the simulation is $0.4/512$ $\mathrm{pc} \sim 8 \times10^{-4}$  $\mathrm{pc}$. Our 3D numerical simulations were carried out across one octant of a sphere.

The refinement algorithm is applied to the surfaces of the contact discontinuity and the two shocks. The surface of the contact discontinuity is detected by means of the gradient in the tracing function $f$ (see Sect.~\ref{eqs1}), rather than the density gradient; in this way, the effectiveness of the detection is independent of the parameters $(n,s)$. The forward and reverse shocks are detected by measuring the pressure discontinuity. The advantage is that the pressure discontinuity is greater than the discontinuities in density or radial velocity and independent of the assumptions about the density power-law index in the free expansion medium and the equation of state of the gas in the interaction region. Figure \ref{ini} shows the angle-averaged radial density profile at $t=50$ yr for $\lambda = 0.57$, or $(n,s) = (7,0)$, and $\gamma = 5/3$ with 1, 2, and 3 levels of refinement. The resolution of the RT structures is crucially improved by applying the AMR. 

During the expansion of a young SNR, the only dynamically interesting region, represented by the interaction region, occupies a fraction of about 40\% (and constant during the self-similar regime) of the whole remnant volume. Moreover the Kelvin-Helmoltz instabilities triggered by the growth of the RT finger-like structures during the ``self-similar'' phase lead to a strongly turbulent phase. Therefore, AMR  hydrodynamic simulation that attempts to spatially resolve the growth of the RT instabilities must refine over an increasingly large relative volume. The contact discontinuity  turns out to be the most computationally expensive region of the flow. In contrast, the core region within the power-law ejecta, where the value of the uniform density is fixed by mass conservation, can be easily handled at the lowest possible refinement level. In this context, the AMR technique, combined with the approach of Moving Computational Grid, is found to be suitable.

We define the boundary conditions of the conservative variables in the corresponding six ghost zones. Since we treat 1/8 of the volume of the remnant sphere, in some ghost zones (three) the boundary condition is inflow, produced by the advection of the interstellar material, and in some other ghost zones (three) the boundary condition is reflexive, according to the location of the interface of the ghost cell. Changes in the order of the assignment of the ghost cells significantly change neither the resulting intercell flow at the shock nor the physical content of the result. In the laboratory rest frame, the ISM is cold and stationary: $\vect{u}_{\rm ISM} = 0$. In the shock frame, this implies that the ISM is advected onto the SNR shock surface with a velocity field given by $\vect{\tilde u}_{\rm ISM} = -a^{\beta} {\dot a}\vect{\tilde r}$ (see Eq.~\ref{newvel}). Therefore, a linear interpolation (second order) for the reconstruction of the velocity field at the boundaries  more clearly describes the inflowing material.

The time step is controlled using a standard Courant factor stability constraint. Based on our expanding coordinate system, we also included a time step limitation, where $a(t)$ is not allowed to vary by more than $10\%$ over the time step. 

The cartesian coordinate system that we use here is not well-adapted to spherically symmetric objects such as SNRs and spurious geometrical effects are expected. It has long been known that an Eulerian code should take care of its violation of Galileian invariance: a turbulent phenomenon observed in a fixed computational grid can disappear if it is observed from a computational grid moving at constant velocity with respect to the fixed grid. The goal of our comoving coordinate system is precisely to eliminate this effect. This improvement does, however, have the problem, which is common to all shock capturing schemes dealing with stationary shocks, of the so-called ``odd-even decoupling'' and the associated carbuncle phenomenon \citep{pi88,q94,pd01}, namely a disturbance in the computation of the flow emerging when the shock direction is aligned with one of the grid coordinate directions. This effect may be amplified when a flow converges in a cell from different directions, as in the case of spherical flow on a cartesian mesh. The result can strongly deform the shock surface and form ripples or bubble-like regions of local sub-density. Local changes to the solver were proposed \citep{kpjm03}, based on an algorithm for the detection of the shock. Reconstruction of the interface values of the state variables is performed, but in the cells detected by that algorithm a more diffusive solver is used, such as HLL, HLLC or Lax-Friedrichs \citep{t99}, while in all the remaining regions a conventional Riemann exact solver or an approximate Roe solver can be used. Solving this long-standing issue is beyond the scope of this paper, so we did not implement these possible fixes here.

\subsection{Integration method}\label{}

We briefly recall the main steps of the integration method of RAMSES \citep{t99}. In the absence of a gravitational potential, the three-dimensional hydrodynamic Euler equations can be written in the conservative form as (see Eq.~\ref{Euler})
\begin{equation}
\frac{\partial {\bf U}}{\partial t} +  \nabla \cdot {\bf F} ( \bf U )  = 0  \, ,
\label{euler system}
\end{equation}
where ${\bf U}$ has components $U_i = (\rho, \rho u_i, E)$, ${\bf F}$ has components $F_{i,j} (U) = (\rho u_{j}, \rho u_{i} u_{j} + \delta_{ij}P, u_{j}(E + P))$, and both $i$ and $j$ have values from $1$ to the space dimension $d=3$. The space-time integration is performed by defining a cell centered on $(x_i, y_j, z_k)$ of volume defined by the coordinates  $V_{i,j,k}=[(x_{i-\half},x_{i+\half}), (y_{j-\half},y_{j+\half}),(z_{k-\half},z_{k+\half})]$ and a time interval  by $\Delta  t = t^{n+1}-t^{n}$, which is not constant through the simulation. In the present section, the quantity $n$ indicates the $n$-th time step of the integration algorithm.
The averaged, cell-centered state is defined by
\begin{equation}
\left< U \right>_{i,j,k} ^n = \frac{1}{\Delta x \Delta y \Delta z} 
\int _{x_{i-\half}} ^{x_{i+\half}} 
\int _{y_{j-\half}} ^{y_{j+\half}}
\int _{z_{k-\half}} ^{z_{k+\half}}
U(x,y,z,t^n)\, {\rm d}x {\rm d}y {\rm d}z\, .
\end{equation}
Apart from a smooth flow of the conservative variables $\left< U \right>_{i,j,k} ^{n+1} $,
the discretization of space imposes the solution to a unidimensional Riemann problem along all the space directions.
Inside the cell $V_{i,j,k}$, the solution $\left< U \right>_{i,j,k} ^{n}$ can be reconstructed with distinct algorithms, for instance constant (first order), linear (second order), to obtain the values at the interfaces
$\left< U \right>_{i-\half,j,k} ^{n}$, 
$\left< U \right>_{i+\half,j,k} ^{n}$,
$\left< U \right>_{i,j-\half,k} ^{n}$,
$\left< U \right>_{i,j+\half,k} ^{n}$,
$\left< U \right>_{i,j,k-\half} ^{n}$, and
$\left< U \right>_{i,j,k+\half} ^{n}$.
The intercell flux is then computed by using the values at the interface.
The time-averaged intercell flux is defined by integrating over the planes separating neighboring cells
\begin{equation}
F_{i+\half,j,k}^{n+\half} = 
\frac{1}{\Delta t \Delta y \Delta z} 
\int _{t^n} ^{t^{n+1}} 
\int _{y_{j-\half}} ^{y_{j+\half}} 
\int _{z_{k-\half}} ^{z_{k+\half}} 
F(x_{i+\half},y,z,t)\, {\rm d}t\, {\rm d}y {\rm d}z\, ,
\end{equation}
\begin{equation}
G_{i,j+\half,k}^{n+\half} = 
\frac{1}{\Delta t \Delta x \Delta z} 
\int _{t^n} ^{t^{n+1}} 
\int _{x_{i-\half}} ^{x_{i+\half}} 
\int _{z_{k-\half}} ^{z_{k+\half}} 
G(x,y_{j+\half},z,t)\, {\rm d}t\, {\rm d}x {\rm d}z\, ,
\end{equation}
\begin{equation}
H_{i,j,k+\half}^{n+\half} = 
\frac{1}{\Delta t \Delta x \Delta y} 
\int _{t^n} ^{t^{n+1}} 
\int _{x_{i-\half}} ^{x_{i+\half}} 
\int _{y_{j-\half}} ^{y_{j+\half}} 
H(x,y,z_{k+\half},t)\, {\rm d}t\, {\rm d}x {\rm d}y\, .
\end{equation}
The conservative system can then be written
\begin{eqnarray}
\left< U \right>_{i,j,k} ^{n+1} - \left< U \right>_{i,j,k} ^{n}  + 
\frac{\Delta t}{\Delta x}
\left(  F _{i+ \half,j,k}^{n+\half} - F _{i-\half,j,k}^{n+\half} \right)  + \nonumber\\
\frac{\Delta t}{\Delta y}
\left(  G _{i,j+ \half,k}^{n+\half} - G _{i,j-\half,k}^{n+\half} \right) +
\frac{\Delta t}{\Delta z}
\left(  H _{i,j,k+ \half}^{n+\half} - H _{i,j,k-\half}^{n+\half} \right) =0 \, .
\end{eqnarray}
This integral form remains exact for the corresponding Euler system since no approximation has been made. 
Since the intercell flux is computed at time $t^{n+\half}$, the method is second order in time. 
The main changes to RAMSES reported here concern the defi\-nition of the inflow of interstellar matter into the simulation box, as explained above, and the introduction of the source term ${\bf S}$. RAMSES allows us to define an analytical acceleration vector, which is treated as an effective gravity vector. We have used this option, the grid motion being defined analytically by Eq.~\ref{at}. The three independent physical quantities, i.e., length, time, and mass, are rescaled according to Eqs.~\ref{Euler1tilde}, \ref{Euler2tilde}, and \ref{Euler3tilde}. All the quantities adopted until the end of the present section are computed in the $\mathbf{\tilde R}$ frame.

In presence of a source term, Eq.~\ref{euler system} becomes
\begin{equation}
\frac{\partial {\bf {\tilde U}}}{\partial {\tilde t}} +  \tilde \nabla \cdot {\bf {\tilde F}} ( \bf {\tilde U} )  = {\bf {\tilde S}}  \, ,
\label{euler system in}
\end{equation}
where ${\bf {\tilde U}}^{n}_{i,j,k}$ denotes a numerical approximation to the cell-averaged
value of $({\tilde \rho}, {\tilde \rho} {\bf  {\tilde u}}_i, \tilde E)$ at time ${\tilde t}^{\,n}$ for the cell $(i,j,k)$,
and the space indices (i,j,k) and the time index (n) are kept for simplicity the same in the $\mathbf{\tilde R}$ frame. 
The numerical discretization of the Euler equations with source terms $S^{n+1/2}_{i,j,k}$ is given by:
\begin{eqnarray}
\frac{{\tilde U}^{n+1}_{i,j,k} - {\tilde U}^{n}_{i,j,k}}{\Delta {\tilde t}} 
& + & \frac{{\tilde F}^{n+1/2}_{i+1/2,j,k} - {\tilde F}^{n+1/2}_{i-1/2,j,k}}{\Delta {\tilde x}} +  \nonumber\\
& + & \frac{{\tilde G}^{n+1/2}_{i,j+1/2,k} -{\tilde G}^{n+1/2}_{i,j-1/2,k}}{\Delta {\tilde y}} +   \nonumber\\
& + & \frac{{\tilde H}^{n+1/2}_{i,j,k+1/2} -{\tilde H}^{n+1/2}_{i,j,k-1/2}}{\Delta {\tilde z}} = S^{n+1/2}_{i,j,k}
\label{euler_source}
\end{eqnarray}
The time centered fluxes ${\tilde F}^{n+1/2}_{i+1/2,j,k}$, ${\tilde G}^{n+1/2}_{i,j+1/2,k}$, 
${\tilde H}^{n+1/2}_{i,j,k+1/2}$ across cell interfaces
are computed using a second-order Godunov method. 
According to Eq.~\ref{Eulertilde}, the gravitational source terms are given by
\begin{eqnarray}
S^{n+1/2}_{i,j,k}  & = &  -\frac{1}{2} \pmatrix{ 0 \cr
{\tilde \rho}^{n}_{i,j,k} (\vect{{\tilde r}}{\mathcal G})^{n}_{i,j,k} + 
{\tilde \rho}^{n+1}_{i,j,k} (\vect{{\tilde r}}{\mathcal G})^{n+1}_{i,j,k}  \cr
({\tilde \rho} {\tilde u})^{n}_{i,j,k} (\vect{{\tilde r}}{\mathcal G})^{n}_{i,j,k} + 
({\tilde \rho} {\tilde u})^{n+1}_{i,j,k} (\vect{{\tilde r}}{\mathcal G})^{n+1}_{i,j,k}  } \nonumber\\
& + & \pmatrix{ 0 \cr 0 \cr {\tilde H}^{n+1} {\tilde \epsilon}^{n+1}_{i,j,k} (2 -  3/\alpha)} \, , 
\label{source}
\end{eqnarray}
where ${\mathcal G}$ is defined in Eq.~\ref{G} and $\tilde \epsilon$ is the internal energy density.
The second term on the right-hand side of Eq.~\ref{source} is equal to zero if $\gamma = 5/3$ or,
equivalently, $\alpha = 3/2$. 
   
Although the integration of the source term $S^{n+1/2}_{i,j,k}$ in Eq.~\ref{euler_source} 
is of second order in time, higher order methods, such as Rosenbrock \citep{kr79}, can be chosen.

\section{Application to SNR}\label{snr}

\subsection{Growth of Rayleigh-Taylor structures}\label{thermal}

\begin{figure*}[!th]
\begin{center}
\includegraphics[width=17.3cm]{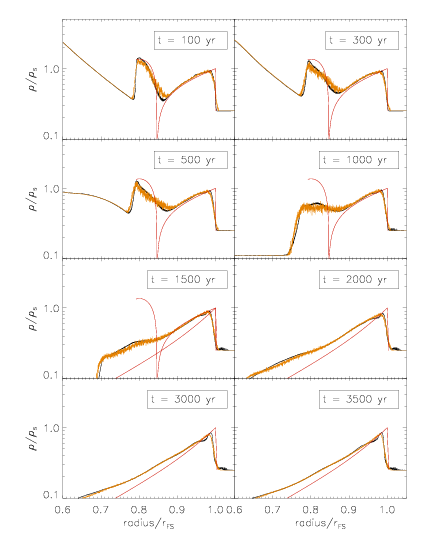}
\caption{Angle-averaged radial density profile at various ages for $\lambda = 0.57$, or $(n,s) = (7,0)$, and $\gamma = 5/3$ with 1 and 3 levels of refinement ($128^3$ in yellow, $128^3-512^3$ in black). For comparison, the red line shows the self-similar profiles normalized to their forward shock values, Chevalier profiles for the first five  panels and Sedov for the last four panels. The departure from the ejecta-dominated phase, occurring when the reverse shock encounters the plateau outer radius $r_{\mathrm c}$, and the transition to Sedov-Taylor phase are shown. The inward motion of the reverse shock, initially only in the comoving frame of the contact discontinuity and later on in the laboratory frame as well, is also depicted. \label{d_1d}}
\end{center}
\end{figure*} 

\begin{figure}
\begin{center}
\includegraphics[width=9cm]{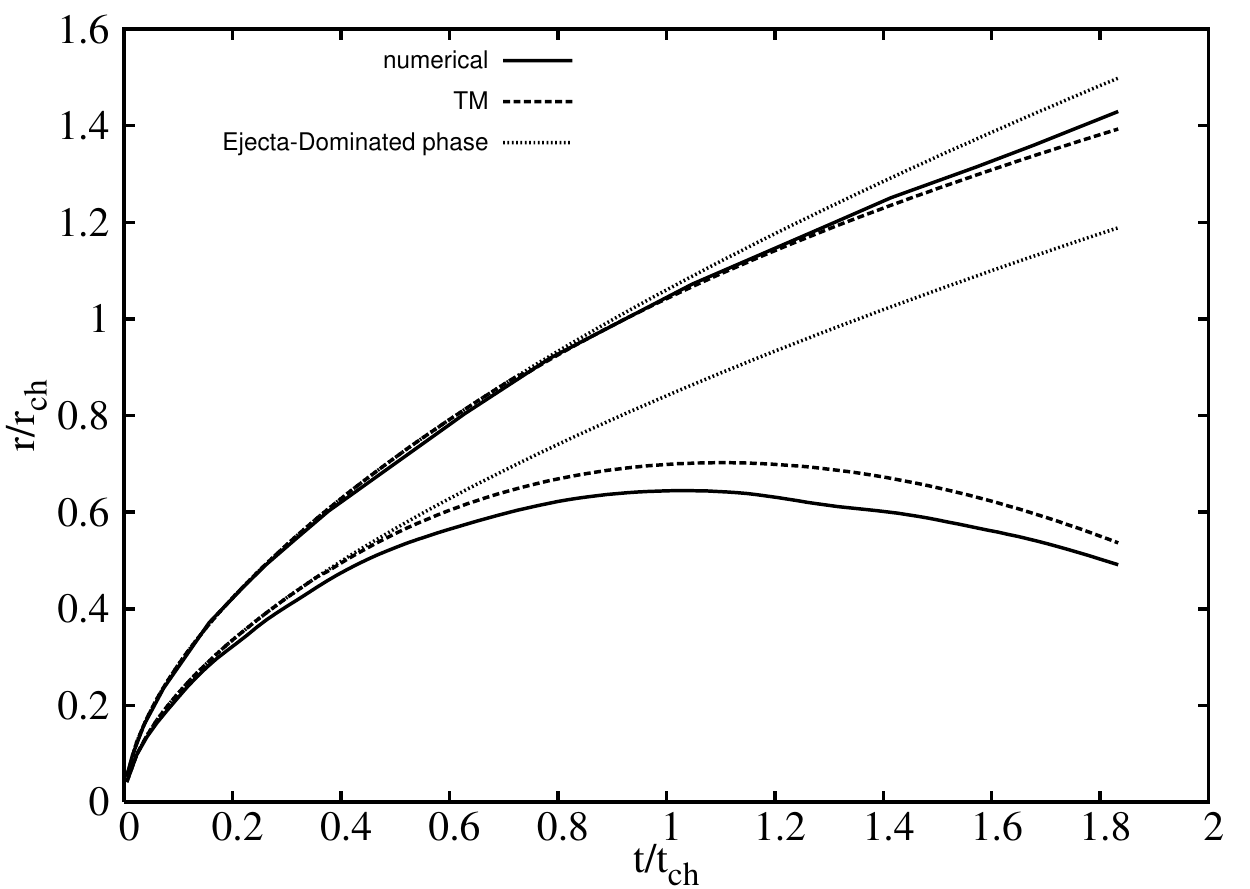}
\caption{Temporal evolution of the forward shock and reverse shock radii for $\lambda = 0.57$, or $(n,s) = (7,0)$, and $\gamma = 5/3$. Our 3D numerical solution ({\it solid}) matches the 1D semi-analytical solution \citep[][]{tm99}, {\it dashed}, and, in the asymptotic regime for $t \rightarrow 0$, matches the self-similar solution in the ejecta-dominated phase \citep{c82}, {\it dotted}. See Sect.~\ref{thermal} for the units $r_{ch}$ and $t_{ch}$ ($r_{ch} \sim 7.8$ pc and $t_{ch} \sim 1950$ yr for the chosen parameters). The spherical symmetry of the reverse shock is strongly modified by the RT instability; here the innermost value of the radius of the reverse shock has been chosen (cf. Sect.~\ref{thermal}). The ejecta-dominated solution (dotted) is prolonged indefinitely to emphasize the inward motion of the reverse shock.\label{r_ch}}
\end{center}
\end{figure} 

\begin{figure*}[!th]
   \centering
   \begin{tabular}{cc}

{ \includegraphics[width=8cm]{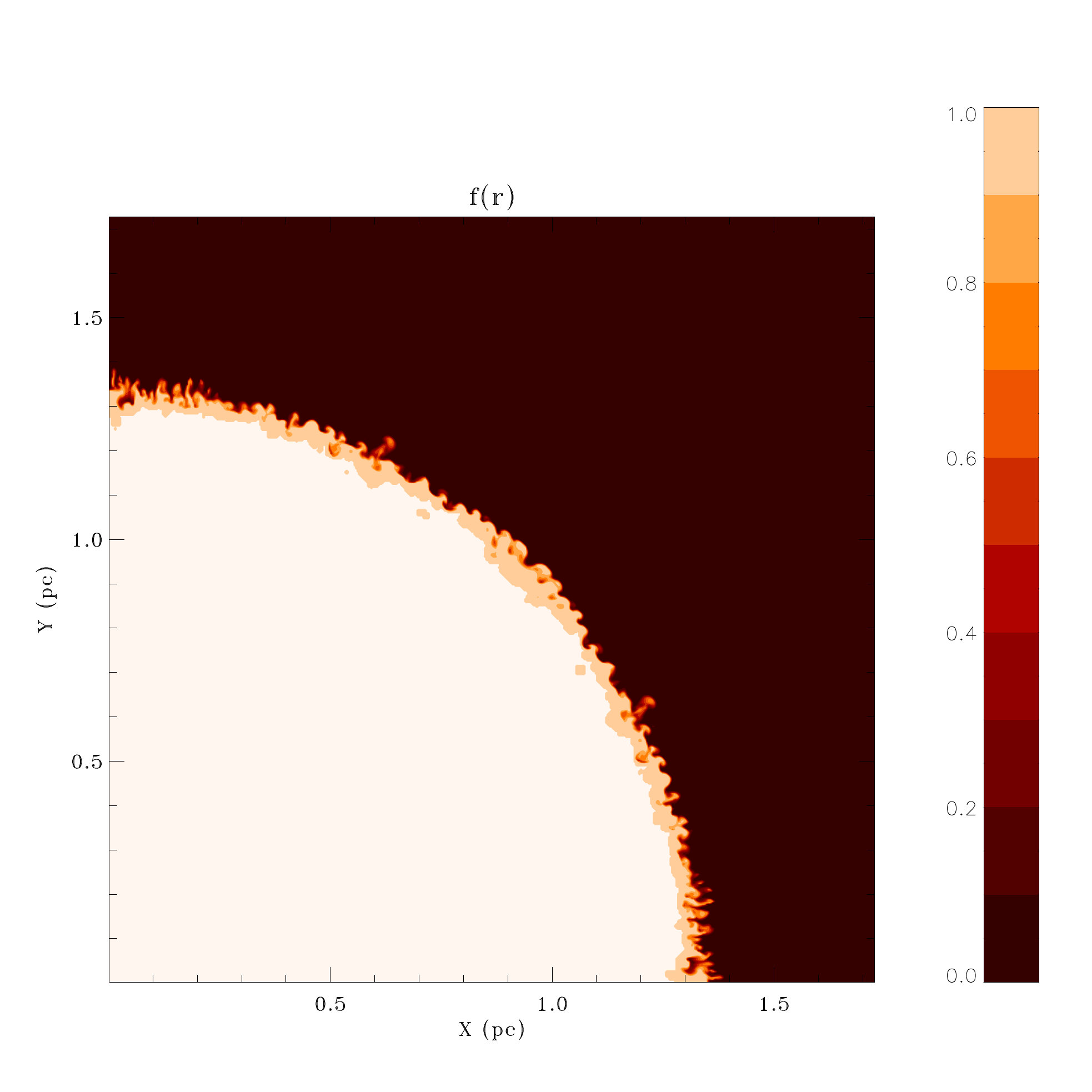} }
{ \includegraphics[width=8cm]{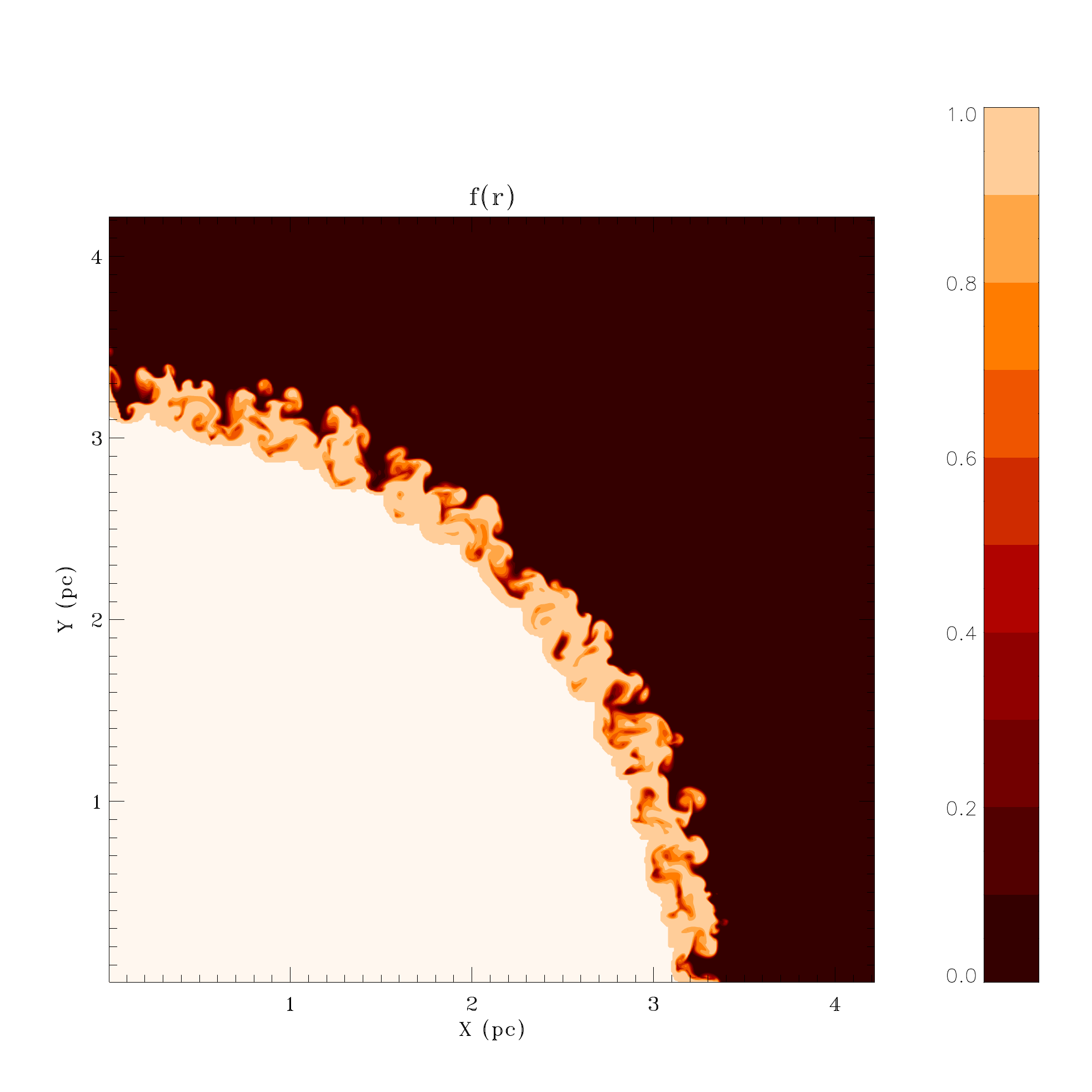} }\\
{ \includegraphics[width=8cm]{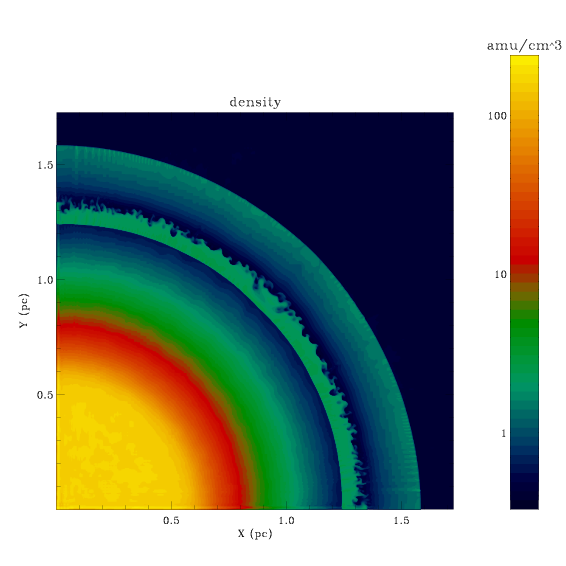} }
{ \includegraphics[width=8cm]{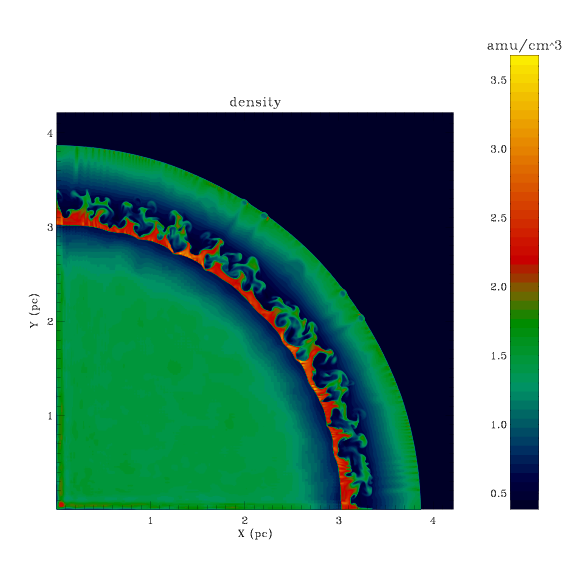}}\\
{ \includegraphics[width=8cm]{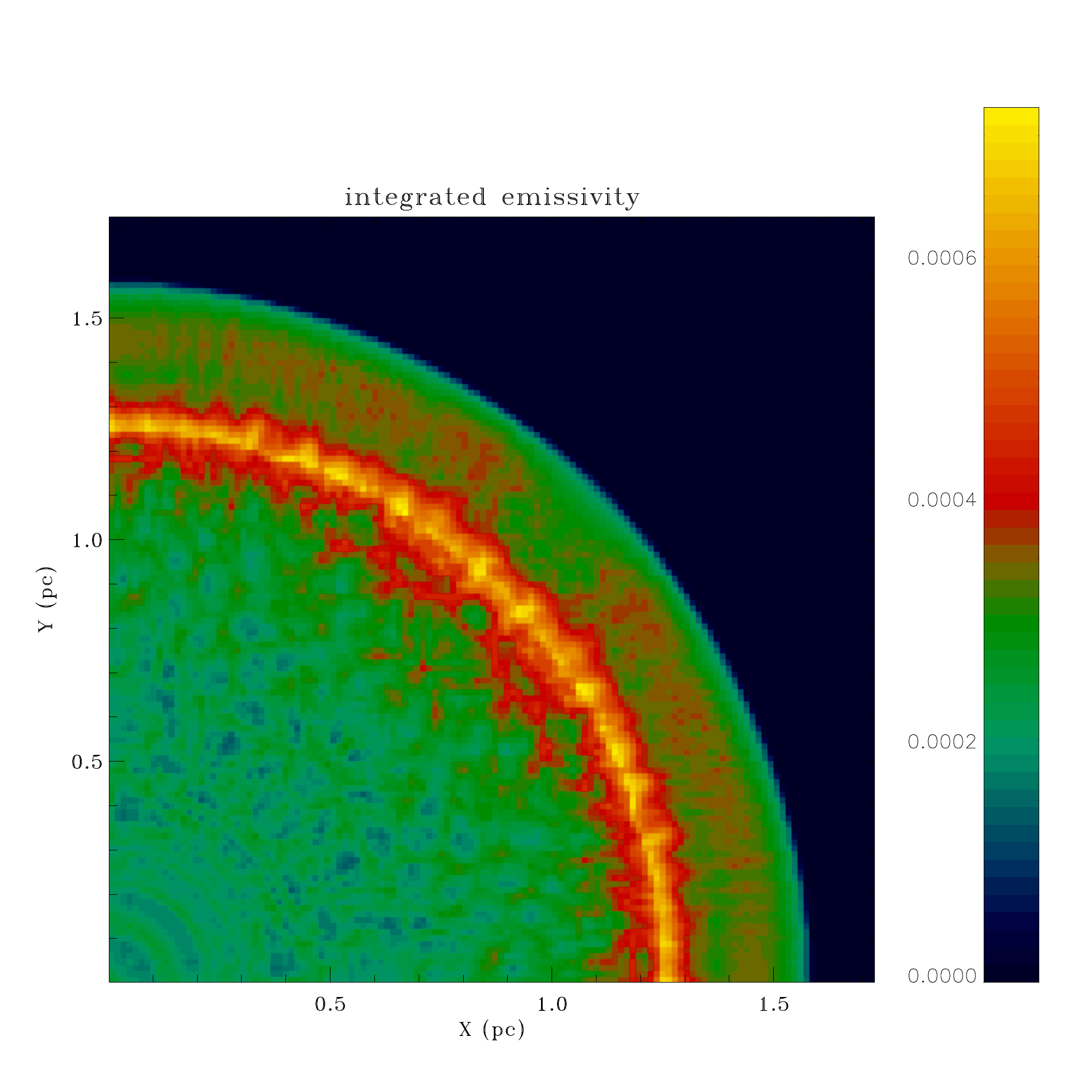} }
{ \includegraphics[width=8cm]{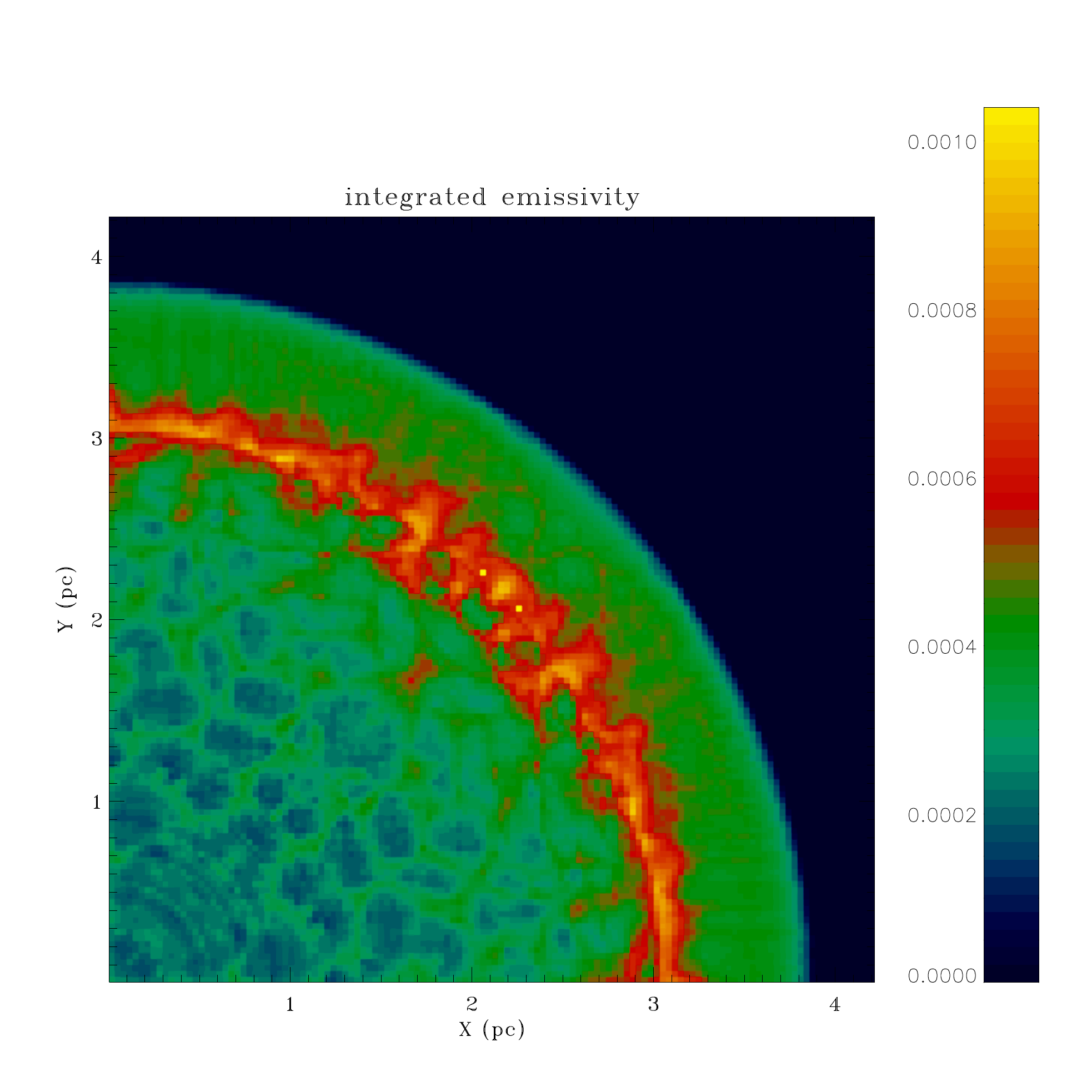} }\\

   \end{tabular}

   \caption{Snapshots at two distinct ages ($t=100$ yr at left; $t=500$ yr at right) for $\lambda = 0.57$, or $(n,s) = (7,0)$ and $\gamma = 5/3$. In the upper panel, the mass fraction $f (\mathbf{r})$, defined in Eq.~\ref{fp}, is shown in a comoving slice plane parallel to the coordinate plane XY at height $z=0.1\times a(t)$. The mixture of the two media, namely ejecta shocked and ISM shocked, as well as the deformation of the contact discontinuity surface is shown. In the middle panel, matter density is shown in the same plane as in the upper panel. Density values are coded according to the color bars given for each frame. Note the change in the radial scale. In the bottom panel, frequency emissivity $J_b$ due to bremsstrahlung integrated along the line of sight towards the reader is shown, assuming a mean molecular weight $\mu = 1/2$, corresponding to the case of fully ionized hydrogen.}

  \label{gamma5}
\end{figure*}

Many developed codes have satisfactorily tackled the growth of RT instabilities as a mandatory validity test. In contrast to previous treatments, we consider the evolution of a full octant of the SNR. Thus, within a larger SNR volume, a large number of RT-instabilitiy lengthscales, spread over a wider range, sum together. In other words, the growth of RT instabilities is more accurately described not only with high spatial resolution, but also with volumes comparable to the global SNR volume. In particular, since the initial departures from the spherical symmetry are sampled over a larger surface, the initial phase of the RT growth is statistically more accurately described (see Sect.~\ref{crd}).
  
Even in the case of a stationary advective shock, a small volume has been dealt with, in the SNR literature, in comparison with the volume of the interaction region. The description here provides a more comprehensive overview of the dynamics of RT instabilities in the SNR.

The simulations presented here cover three phases of the adiabatic SNR expansion (see Fig.~\ref{d_1d}): the self-similar expansion \citep{c82} with an exponent $\lambda$ depending on the properties of the ejecta and the ISM, i.e. on $(n,s)$; the second phase, not self-similar, representing the transition to the Sedov-Taylor regime ($M_{\rm ISM} \sim M_{\rm ej}$); and the third phase (Sedov-Taylor), with a blast wave radius $R_b$ that expands as $R_b \propto t^{2/5}$,  for $s=0${\bf ,} and $M_{\rm ISM} \gg M_{\rm ej}$. The transition between the two self-similar phases is shown by means of various snapshots of the angle-averaged density in Fig.~\ref{d_1d}. In this case, we assume that $\lambda = 0.57$, or $(n,s) = (7,0)$, and $\gamma = 5/3$ in the whole simulation box. The inward motion of the reverse shock and the progressive establishment of the Sedov-Taylor profile is manifest. We extended our computation until values of time $t$ when the inward motion of $ r_{\mathrm{RS}}$ has reached the inner core, and show that  during the transition to the Sedov-Taylor phase, the angle-averaged profile is not significantly altered by the deve\-lopment of the RT instability. The 1D Sedov-Taylor density profile is superimposed in the last three panels of Fig.~\ref{d_1d}, by assuming that the Sedov-Taylor phase is fully established at $t = 5 \times t_{\mathrm{ST}}$, where $t_{\mathrm{ST}} = 0.732  \times t_{ch}$ \citep{tm99}. Compared with the original version of the code in the case of a spherical Sedov blast wave \citep{t02}, we note the improvement in the resolution of the compression factor; in that previous work relatively small values of final output times had been considered, the aim being to test the code.

\begin{figure}
\includegraphics[width=8.5cm]{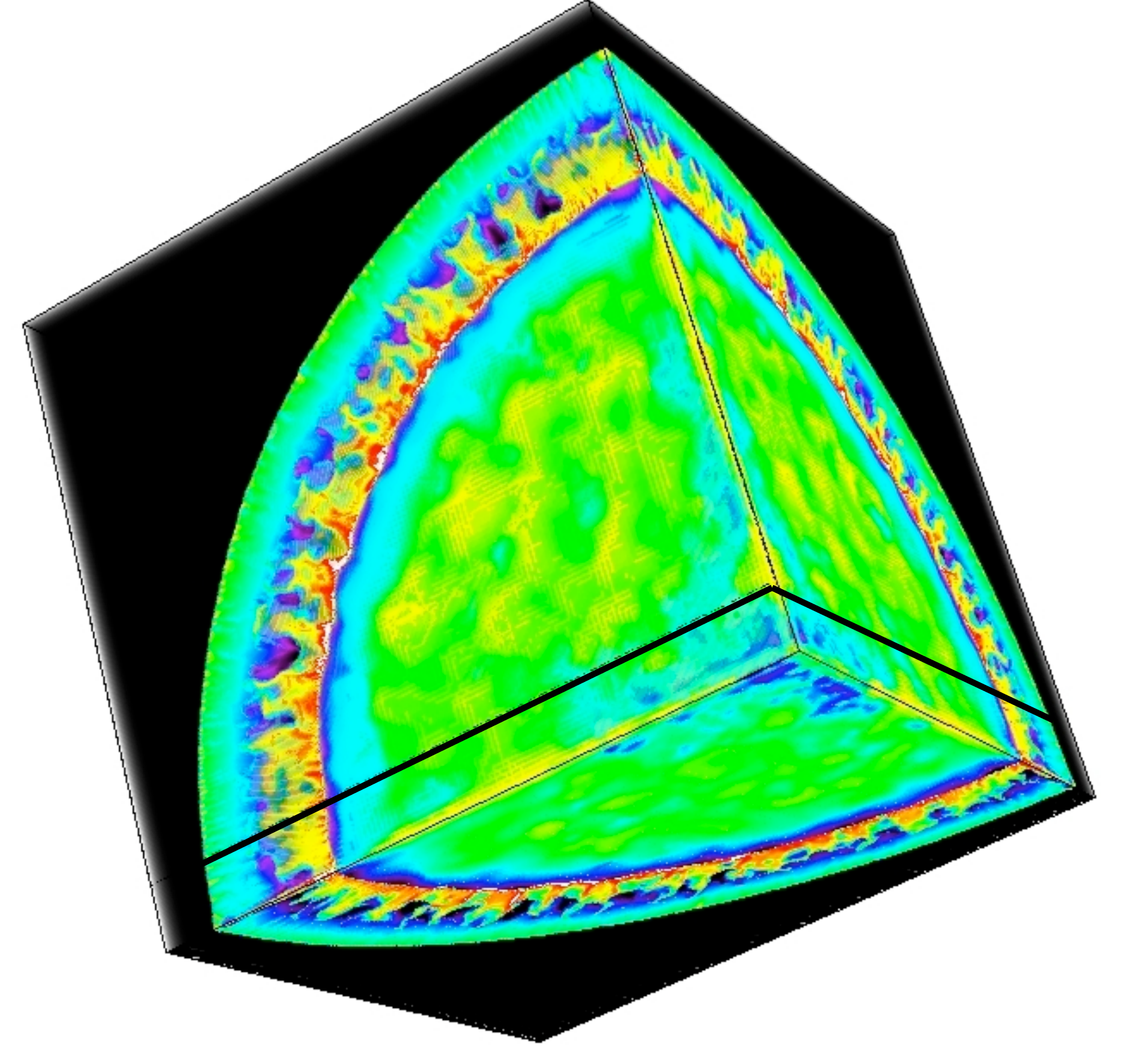}
\caption{The three-dimensional density in the full simulation box of side $L = 4.2$ pc is shown at $t=500$ yr for $\lambda = 0.57$, or $(n,s) = (7,0)$ and $\gamma = 5/3$. The black line indicates the height of the planar slice at $z = 0.1\times L$ used in Fig.~\ref{gamma5}, upper and middle panels.}
\label{density3D}
\end{figure}

In Fig.~\ref{r_ch}, the outermost value of the position of the forward shock and the innermost value of the position of the reverse shock in 3D are compared with the semi-analytical 1D solution of \citet{tm99}. The units used here correspond to $r_{ch}$ (pc) $= 3.07 (M_{\rm ej}/M_{\odot})^{1/3} (\rho_{\rm ISM}/\mu_{\rm H})^{-1/3}$ and $t_{ch}$ (yr) $= 423 E_{51}^{-1/2} (M_{\rm ej}/M_{\odot})^{5/6} (\rho_{\rm ISM}/\mu_{\rm H})^{-1/3}$, where $\mu_{\rm H}=2.34\times10^{-24} g$ is the mean mass per hydrogen nucleus assuming cosmic abundances. In particular, the inward motion of the reverse shock is compared with the solution of \citet{tm99}; in the case $(n, s) = (7, 0)$, we find agreement within $7\%$ between our 3D simulation and the 1D solution of \citet{tm99}. The radius of the reverse shock is strongly modified by the RT instability where the innermost value is chosen. In contrast, the radius of the forward shock is unequivocally defined, since it is not perturbed by the RT instabilities for the physically relevant values of $(n,s)$ (see also Fig.~\ref{density3D}).

In Fig.~\ref{gamma5} (upper panel), the mass fraction $f(\mathbf{r})$, defined in Eq.~\ref{fp}, is shown in a planar slice parallel to the coordinate plane XY at time $t=100$ yr (left) and $t=500$ yr (right); in this case, $\lambda = 0.57$, or $(n,s) = (7,0)$ and $\gamma = 5/3$. The mixture of the two media, shocked ejecta and shocked ISM, as well as the deformation of the contact discontinuity surface appear at intermediate values of $f(\mathbf{r})$. In the middle panel, the matter density is shown with the same spatial extension and times as in the upper panel. In the lower panel, an estimate of the frequency emissivity $J_{\rm b}$ produced by bremsstrahlung in purely hydrogen gas is shown, integrated along the line of sight towards the reader. Only the interaction region  where the temperature is high contributes to the emissivity. The emissivity $J_{\rm b}$ is computed based on the assumption that the plasma is optically thin (bremsstrahlung). Therefore, the emissivity is given by $J_{\rm b} = 2\times10^{-27} Z^2 {N_{\rm e}}^2 T_{\rm e} ^{1/2}$, where $N_{\rm e}$ is the electron number density and $T_{\rm e}$ is the electron temperature given by the perfect gas equation of state such that $T_{\rm e} = \mu m_{\rm H} P/(\rho {\rm k})$, where $\mu$ is the mean molecular weight, $m_{\rm H} = 1.67\times 10^{-24} $g is the proton mass, and ${\rm k} = 1.3807\times 10^{-16}$ erg/K is the Boltzmann constant. For simplicity, in this case we chose $\mu = 1/2$, corresponding to the case of fully ionized hydrogen. We implicitly assumed that $T_{\rm e} = T_{\rm i}$, where $T_{\rm i}$ is the ion temperature, while at an 
early stage the SNR is not yet completely thermalized. A more detailed study should account for inclusion of the evolution of $T_{\rm e} /T_{\rm i}$, but this indicative diagram provides nevertheless a starting point.

In Fig.~\ref{density3D}, a 3D large view image of the density field in the full simulation box is shown. This picture shows that while the sphericity of the forward shock surface is maintained since the RT instabilities do not reach the shock, the reverse shock surface is strongly modified. The large relative volume of the refinement region can be clearly seen.

\begin{figure*}[!th]
   \centering
   \begin{tabular}{cc}

{ \includegraphics[width=8cm]{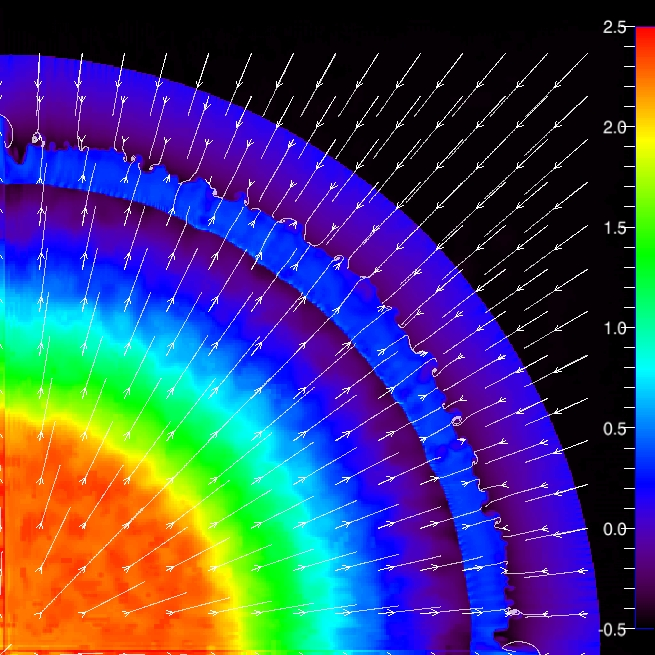}}
{\includegraphics[width=8cm]{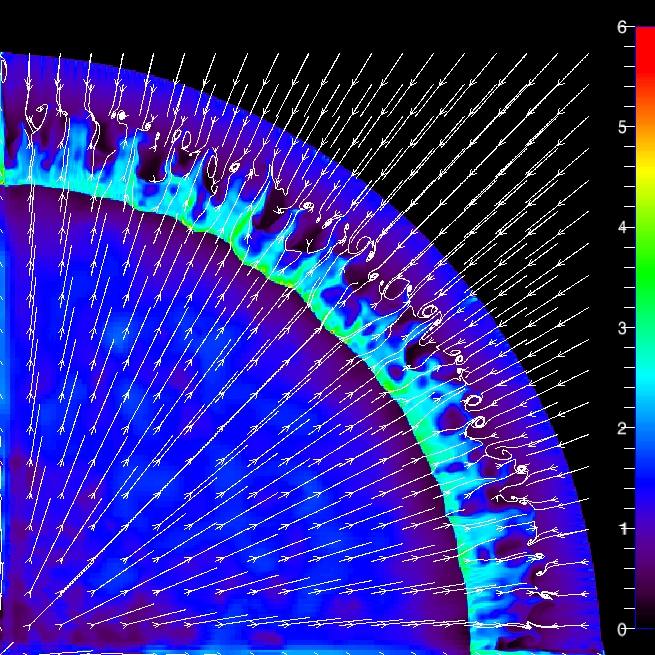}}\\
{ \includegraphics[width=8cm]{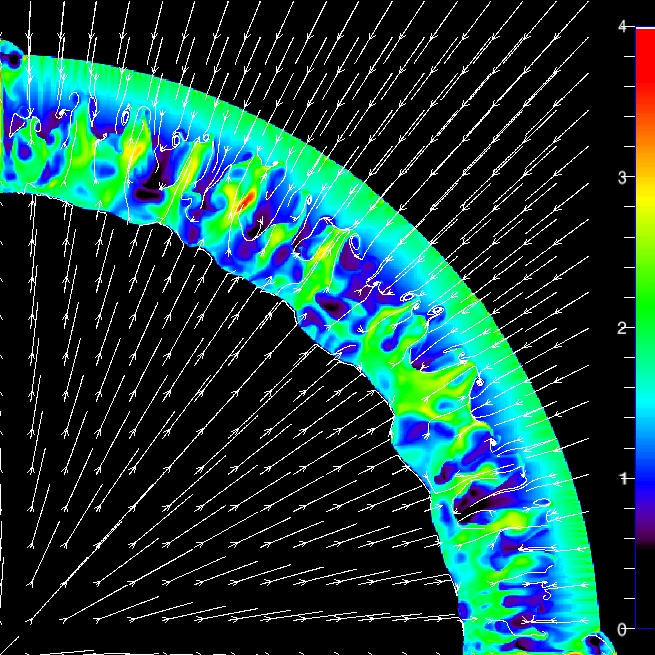} }
{ \includegraphics[width=8cm]{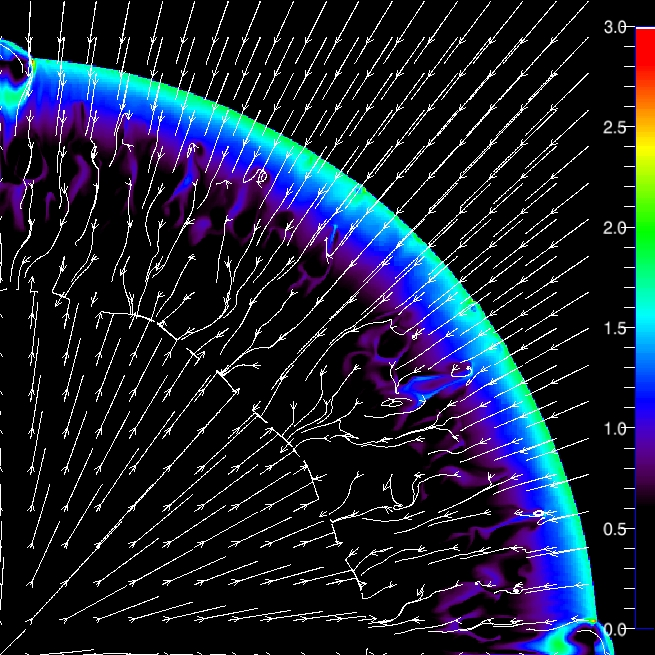}}\\
   \end{tabular}

   \caption{Snapshots showing streamlines of velocity field superposed to matter density for $\lambda = 0.57$, or $(n,s) = (7,0)$ and $\gamma = 5/3$ in the frame of contact discontinuity. The planar slice is in all panels z=0. From left to right, in above panel ($t=100$ yr) the inflow motion of the ISM and the outflow motion of the freely-expanding ejecta are clearly evident. In the other panels ($t=500$ yr, $t=1000$ yr, $t=2000$ yr), the turbulent flow motion due to RT instabilities is shown. Note that the scale in color bar is log in amu$\times$cm$^{-3}$ in the first panel and linear in all the other panels.}

   \label{vel2d}
\end{figure*}

In Fig.~\ref{vel2d}, the streamlines of the velocity field in the reference frame of the contact discontinuity in distinct snapshots show the development of the chaotic flow in the interaction region due to the convective motions. As the time proceeds, the central density decreases and the relative size of the region of turbulence increases. 

\begin{figure}
\includegraphics[width=9cm]{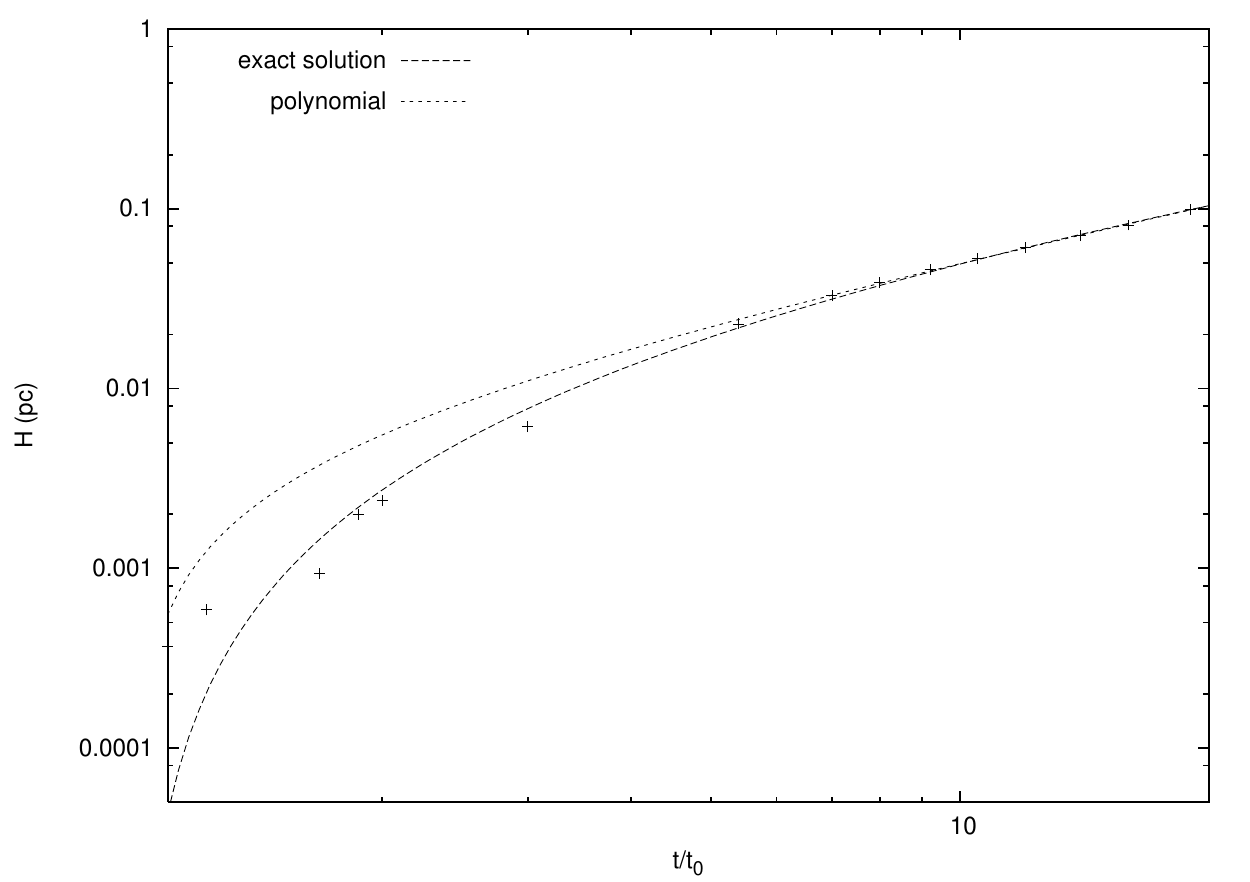}
\caption{The protruding RT structures measured in our simulations at different times are quite well reproduced by the exact solution $H(t)$ with $\lambda = 0.57$ (see Eq.~\ref{SolEx}). The edge of the RT structures is defined by $f(r) \sim 0.01$. The early-time exponential growth can hardly be verified with the available spatial resolution. A second order polynomial fit is also shown for comparison.}
\label{Ht}
\end{figure}

For a general acceleration field $g$, the initial growth of small amplitude perturbations at the accelerated contact discontinuity between two incompressible fluids is exponential with a growth rate $\sigma$. The quantity $\sigma$ is related to the wavenumber $k$ by the relation $\sigma = \sqrt{kgA}$, where $A = (\rho_1 - \rho_2)/(\rho_1 + \rho_2)$ and $\rho_1$, $\rho_2$ are the densities of the two adjacent media at the contact discontinuity. At later times, the growth of the ripples at the contact discontinuity enters the so-called ``self-similar phase''; this phase was first quantitatively analyzed by \citet{fv53}. The self-similar growth of the RT instable structures is governed by the equation
\begin{equation}
\frac{dH(t)}{dt} = 2 (\delta A g)^{1/2} H^{1/2}(t) \, ,
\label{RTgrowth}
\end{equation}
where the constant $\delta$ is a dimensionless growth parameter depending on the dimensionality and the geometry of the system, and $g$ is the acceleration. If $g = const$, the asymptotic quadratic law in time is found to be $H_1(t) = \delta A g (t - t_0)^2$. A second order polynomial fit was found to reproduce quite satisfactorily the mixing of ejecta and ISM in a young SNR \citep{d00}. However, in view of the self-similar deceleration of the contact discontinuity of an SNR, it is reasonable to assume a time-dependent $g$. We propose that a more accurate result can be found by assuming that the acceleration $g$ is also self-similar, namely $g(t) \sim t^{\lambda - 2}$. The integration of Eq.~\ref{RTgrowth} gives readily
\begin{equation}
\frac{H(t)}{r_{\rm CD}(t_0)} = \frac{4\delta A (\lambda-1)}{\lambda}\left[\left(\frac{t}{t_0}\right)^{\lambda/2} - 1\right]^2 \, ,
\label{SolEx}
\end{equation}
where $r_{\rm CD}(t_0)$ is the initial radius of the contact discontinuity. The solution in Eq.~\ref{SolEx} reproduces our simulation (see  Fig.~\ref{Ht}). The comparison in Fig.~\ref{Ht} shows a disagreement of a factor of more than $2$ with the second order polynomial fit at early times, because the deceleration of the contact discontinuity has not been properly taken into account.

The limited spatial resolution attainable at the contact discontinuity with a cartesian grid makes it hard to verify early time exponential growth \citep{fv53}, prior to the self-similar regime.

\subsection{Non-thermal particles in the shocked ISM}\label{nonthermal}

Multiwavelength surveys have inferred that cosmic-ray acceleration can be efficient in young SNRs \citep[see][and references therein]{v08b}. The idea at the basis of particle acceleration at SNR shocks is the Fermi first-order mechanism of acceleration,
namely that particles accelerate by repeatedly crossing the shock and colliding at every passage with magnetic turbulences. 
Therefore, beside other parameters, the efficiency of acceleration depends on the intensity and the spatial configuration of the magnetic field. The X-ray detections of narrow synchrotron emitting filaments at SNR shocks indicate magnetic field intensities  up to $1$ mGauss \citep{b03,l03,vl03}. Magnetic field amplification at shocks has been interpreted as the result of turbulence in the postshock material \citep{gj07} or cosmic-ray induced streaming instability \citep{b04,ab09}.

The SNR hydrodynamics can be modified by efficient particle acceleration at the forward shock \citep{deb00}.
Therefore, a complete numerical description of the SNR expansion should take into account the feedback of the non-thermal population of particles on the dynamics of the SNR, i.e., the system of Euler equations (Eqs.~\ref{Euler}) should be coupled 
to the equation for the cosmic-ray phase-space distribution function.

\citet{be01} investigated this scenario by changing the adiabatic index $\gamma$ globally.
They found that the development of the RT instability was relatively
unaffected and the main effect was that, because the compression ratio
increases and the shocked region shrinks when $\gamma$ decreases, the RT
fingers approached the forward shock and could even reach it for $\gamma = 1.1$.
We adopt a similar approach but investigate when particle acceleration is efficient at the forward shock only. 
We call this the hybrid case in the remainder of the paper.
A local equation of state of the form $P = (\gamma(r) -1) \epsilon$ is assumed. The shocked ISM is initialized as a relativistic gas 
(dominated by cosmic rays) and the shocked ejecta is initialized as a non-relativistic gas such that
\begin{equation}
\gamma(r) = \left\{
  \begin{array}{cc}
    5/3 & \rm{if~} r < r_{CD} \\
    4/3 & \rm{if~} r > r_{CD}  \, ,
    \label{gamma}
   \end{array}
\right.
\end{equation}
the corresponding $\alpha$ then being propagated according to Eq.~\ref{Eulertilde}.

\begin{figure}
\centering
\includegraphics[width=9.3cm]{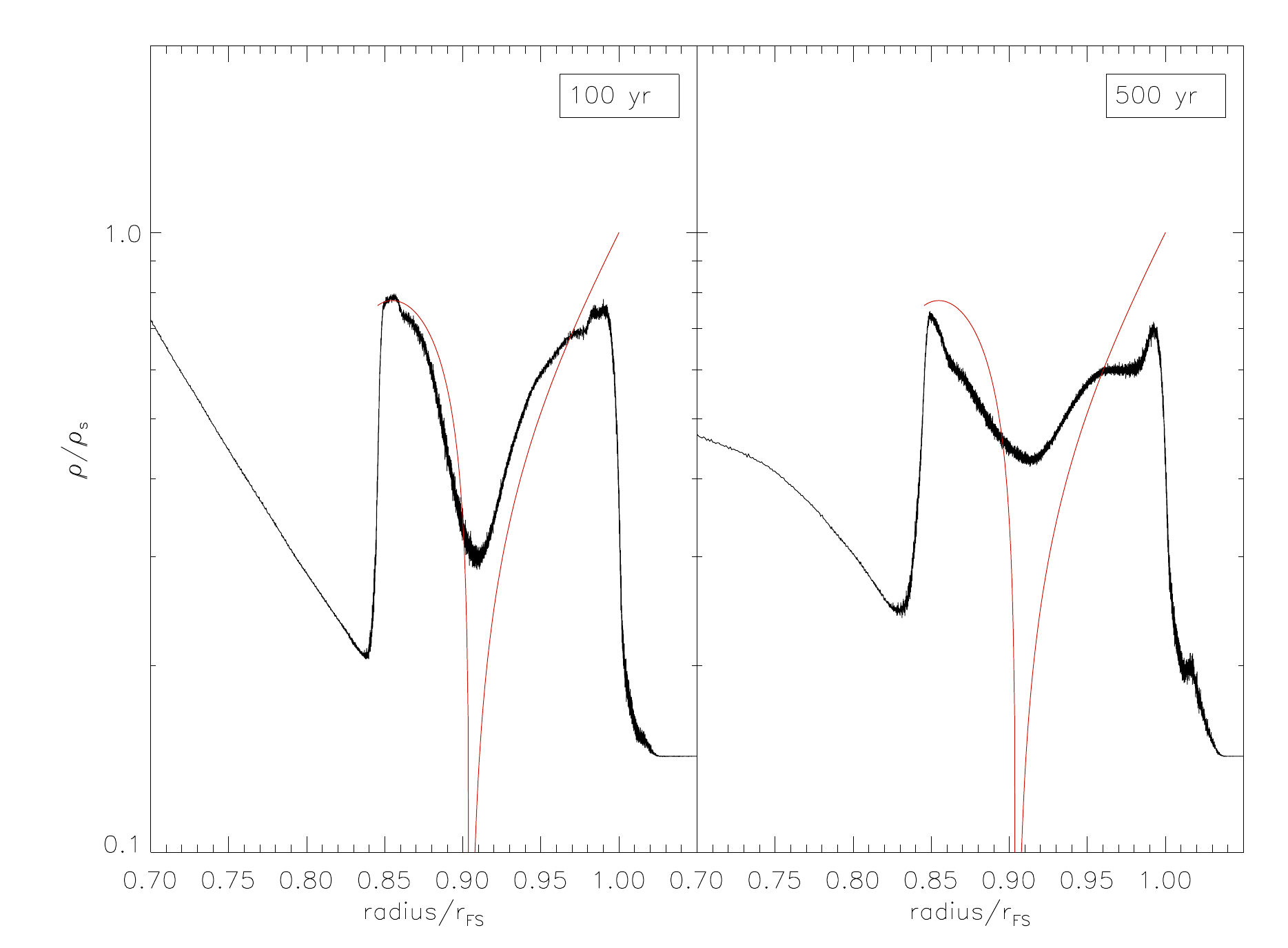}
\caption{Angle-averaged radial density profile shown at distinct ages for $\lambda = 0.57$, or $(n,s) = (7,0)$ in the hybrid case in black, is compared with the corresponding initial profile, in red.}
\label{d_comp3}
\end{figure}

The initial adiabatic $\gamma$ distribution in Eq.~\ref{gamma} should provide a more realistic representation of forward shock accelerating particles than a $\gamma = 4/3$ uniform distribution across the whole interaction region.
There has been relatively little observational or theoretical evidence that particle acceleration is very efficient
at the reverse shock \citep{edb05}. An interesting result of that simulation is that the density behind the reverse shock is lower
than behind the forward shock in the self-similar solution, at least for $n = 7$ (Fig.~\ref{d_comp3} as opposed to Fig.~\ref{ini}).
This may hinder the development of the RT fingers.

\section{Discussion}\label{disc}

\subsection{Intrinsic issues in the stationary advective shock problem}

In simulating stationary advective shocks, numerical instabilities related to the ``carbuncle phenomenon'' can be found by using the Roe solver. The carbuncle phenomenon was first observed by \citet{pi88} in blunt body simulations using Roe's method and numerically studied in detail by \citet{q94}, and has remained a crucial numerical issue \citep{pd01,xl01} until recently \citep{nk08}. The carbuncle phenomenon is considered to be a spurious numerical effect affecting hydrodynamic simulations when the intercell flow is aligned with one of the coordinate directions of the computational grid; it shows up when a supersonic shock, or more generally a discontinuity in the flow variables, is comoving with the computational grid. The Riemann solver produces unphysical ripples of the shock and in the post-shock region, which at late times can modify the structure of the shock surface itself. The numerical methods that are likely to produce carbuncle phenomenon have not yet been unequivocally identified, an attempt to classify them being \citet{dmg03}.

The so-called ``H-correction'' \citep{sgths08}, which consists of introducing an anisotropic dissipative flux in the direction perpendicular to the direction of propagation of the shock, cannot be applied here. We did not add a diffusive term in the intercell flux, which is usually adopted to avoid the growth of numerical instabilities. The addition of artificial viscosity could damp physical effects emerging during the development of the RT instabilities, which we are interested in studying in detail. For the same reason, we did not use more diffusive solvers such as HLL or HLLC or Lax-Friedrichs \citep{t99}. 

In the case of supernova core-collapse simulations, a possible solution has been to compute the intercell flux in the zones of strong shock with an Einfeldt solver \citep{q94,q97}, while all the other grids in the computational box are treated with a less diffusive Riemann solver \citep{kpjm03}. We preferred not to introduce a solver-switching algorithm to avoid introducing other spurious effects possibly arising from the matching of the intercell fluxes. 

To avoid the numerical instability occurring at the stationary shock, we used the exact solver. The failure of the Roe solver manifests itself in local fluctuations of density in regions behind the shock and strong deformations of the  shock surface, which grow as the non-linear regime of RT instabilities is reached; consequent unphysical effects due to velocity shear, resembling Kelvin-Helmholtz instabilities, develop at the same time. Both effects are absent if the shock is not stationary with respect to the computational grid and depend on the chosen simulation parameters, for instance CFL number or interpolation scheme of conservative variables; thus they must be attributed to the choice of the solver. The choice of the exact solver did not significantly increase the computational time.

\subsection{Quasi-comoving grid}

A different solution to the carbuncle problem could involve defining a computational grid that drifts with respect to the SNR, slowly enough to follow the expansion of the remnant for a long time and to avoid boundary reflections possibly occurring with a remnant overtaking the grid. However, this is an even less viable solution since, as shown in the following, the angle-averaged density profiles strongly deviate from the expected compression ratio at the two shocks. In contrast, as shown earlier, in the case of an exactly comoving computational grid, the post-shock ripples average to zero.

Independently of the initial density profile, i.e., of the value of $\lambda$, the self-similar regime slows down towards the Sedov-Taylor solution, which has a self-similar exponent equal to $2/5 <\lambda$. Therefore, the contact discontinuity in the computational grid will progressively acquire a drift velocity depending on the physi\-cal parameters of the remnant. To test the numerical stability of the shock, we introduced a ``quasi-comoving'' grid, whose small drift with respect to the contact discontinuity velocity is parametrized by $\varepsilon \ll 1$: $a(t) = a_0 (t/t_0)^{\lambda + \varepsilon}$.

\begin{figure*}[!th]
  
   \begin{tabular}{cc}

{ \includegraphics[width=8.9cm]{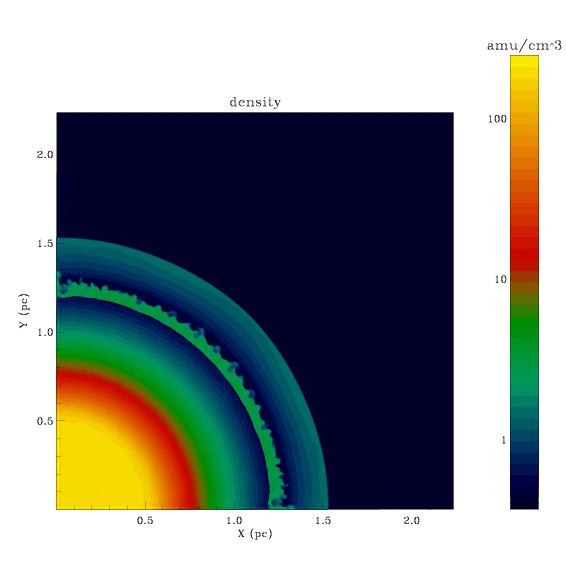} }
{ \includegraphics[width=8.9cm]{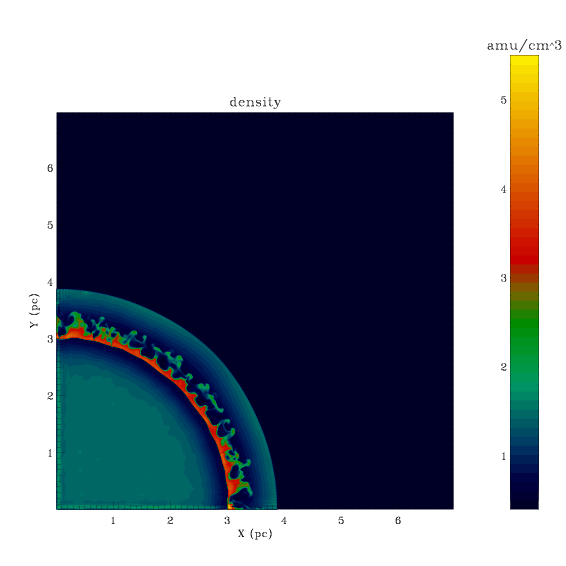}}\\
   \end{tabular}

   \caption{Matter density in a comoving planar slice parallel to the coordinate plane XY at height $z=0.1\times a(t)$ ($t=100$ yr at left; $t=500$ yr at right) for $\lambda = 0.57$, or $n=7$, $s=0$, and $\gamma = 5/3$. Density values are coded according to the color bars given for each frame. Note the change in the radial scale. In this case, $ \varepsilon = 0.129$. Since $\lambda' > \lambda$, the computational grid expands faster than the SNR and the volume fraction occupied in the simulation box by the SNR decreases. It is evident that even a small drift of the computational grid from the shock velocity erases the effects of the numerical instability possibly emerging behind the shocks.}

   \label{gamma6}
\end{figure*}

\begin{figure}
\includegraphics[width=9.6cm]{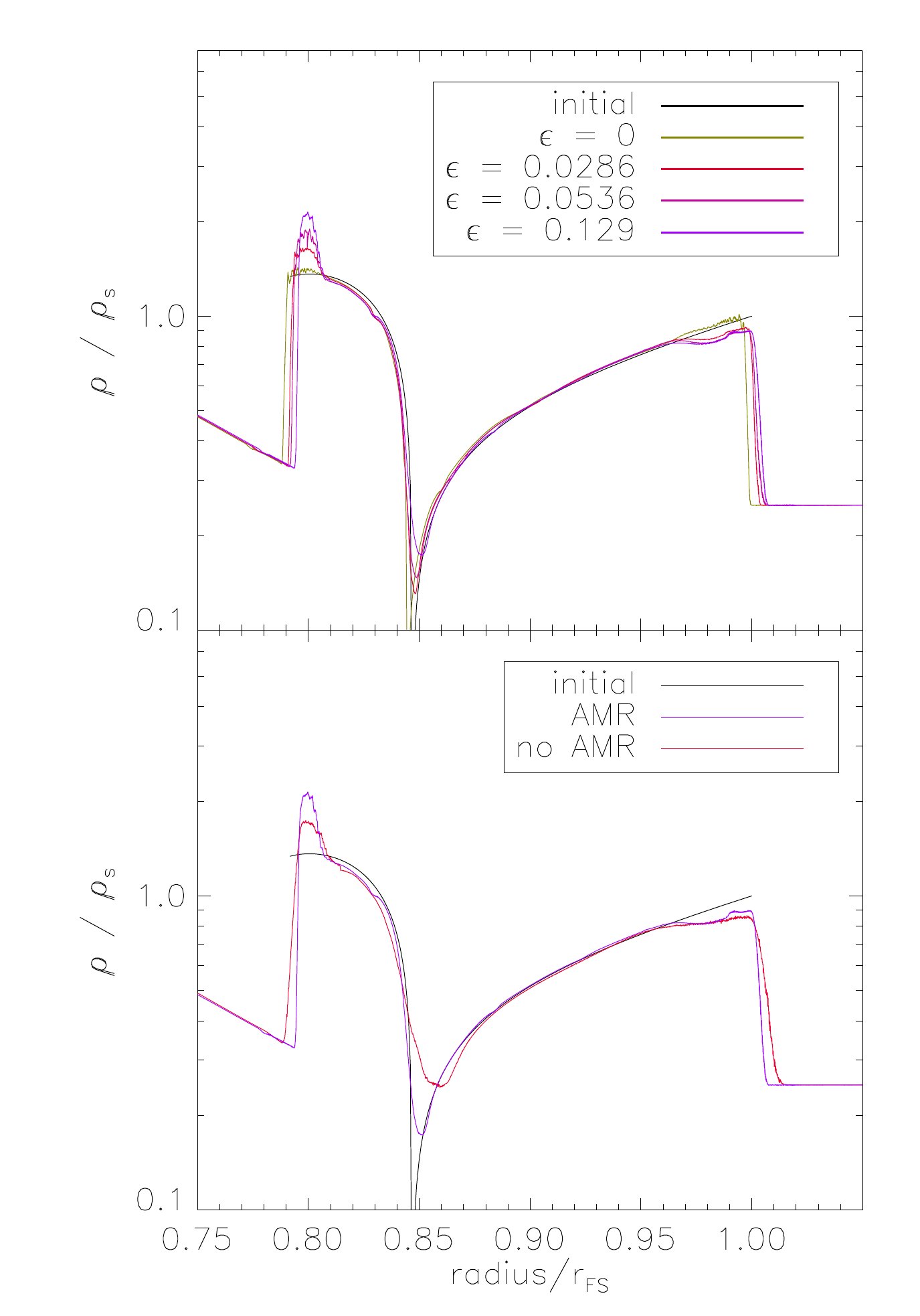}\\
\caption{Angle-averaged radial density profiles between the reverse shock and the forward shock at time $t=12$ yr = 0.0063 $  t_{ch}$, with $\lambda = 0.57$, or ($n,s)=(7,0)$, and $\gamma = 5/3$. The AMR is performed over three levels of refinement. The radial coordinate is normalized to the forward shock position corresponding to $\varepsilon = 0$; the density is normalized to the value at the shock. {\it Upper panel.} The profile in the case of the exactly comoving grid $(\varepsilon = 0)$ is compared with the profiles having a quasi-comoving grid $\varepsilon = (0.0286, 0.0536, 0.129)$. The initial profile, at $t_0 = 10$ yr, is also depicted for comparison. It is evident that even small values of $\varepsilon$ produce oscillations and an error of the order of $10\%$ in the compression factor for $\varepsilon = 0.129$. Outside the interaction region, the profiles are superimposed. {\it Lower panel.} The curves without AMR and with AMR are compared with the initial profile at $t_0=10$ yr. In this case, $\varepsilon = 0.129$. The independence of the error in the compression factor at the two shocks from the space resolution is clear.}
\label{quasi}
\end{figure}

The quasi-comoving grid, even with the Roe solver, smears out the ripples produced behind the forward shock surface by the stationarity of the computational grid, therefore locally is odd-even instability-free (see Fig.~\ref{gamma6}). However, the angle-averaged density profile, even at the first time-steps, is affected by an error in the compression factor at the two shocks (see Fig.~\ref{quasi}, upper panel), the position of the shock being unchanged. The redistribution of the mass through the two shocks in the  presence of a grid drift (over-compression at reverse shock and under-compression at forward shock) is to be attributed  to the mapping of a spherically symmetric object on a cartesian grid. It manifests itself during the transition to the Sedov-Taylor phase in terms of an error of $5\%$ in the resolution of the compression at the shock, and should be considered an intrinsic limit of the approach used here. This error does not depend on the number of levels of the AMR (see Fig.~\ref{quasi}, lower panel). As a consequence, even if a larger number of computational cells is considered, the angle-averaged value at the shock does not fulfill the value of the compression predicted by the Rankine-Hugoniot conditions. In contrast, for an exactly comoving grid, even with the Roe solver, strong deformations of the shock surface are triggered by the numerical instability but the compression at the shocks is preserved. The same difference of results between the exactly comoving and quasi-comoving has been found with the exact Riemann solver, shown in Fig.~\ref{quasi}. 

As a consequence, during the transition to the Sedov-Taylor solution, the generation of a numerical error must be expected in the compression at the shock. A possible solution is the introduction of a grid velocity given by $a(t) = a_0 (t/t_0)^{\lambda(t)}$.

\subsection{Cosmic-ray-dominated blast wave}\label{crd}

\begin{figure}
\centering
\includegraphics[width=9.2cm]{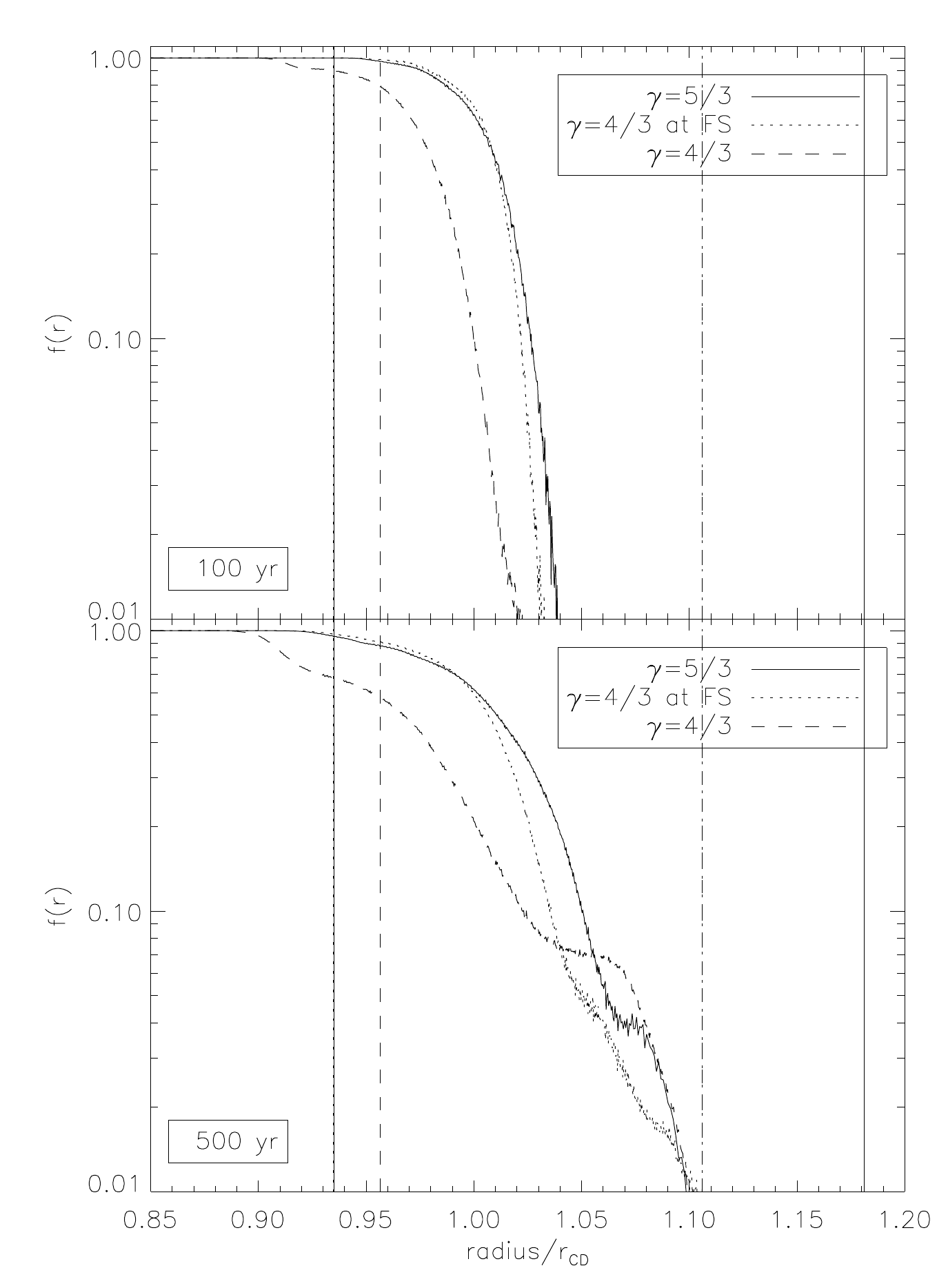}
\caption{Angle-averaged radial profile of the mass fraction of the ejecta $f$ shown at distinct ages for $\lambda = 0.57$, or $(n,s) = (7,0)$; the cases with uniform $\gamma = 5/3$, $\gamma = 4/3$ and the hybrid case are compared. The radius is in units of the corresponding radius of the contact discontinuity. The vertical lines on the right of the panels indicate the corresponding positions of the forward shock, which at this age still meets the self-similar expansion (see Fig.~\ref{r_ch} for the uniform case). The vertical line on the left of the panels indicates the position of the reverse shock. }
 \label{fp_comp2}
\end{figure}

\begin{figure}
\centering
\includegraphics[width=9.2cm]{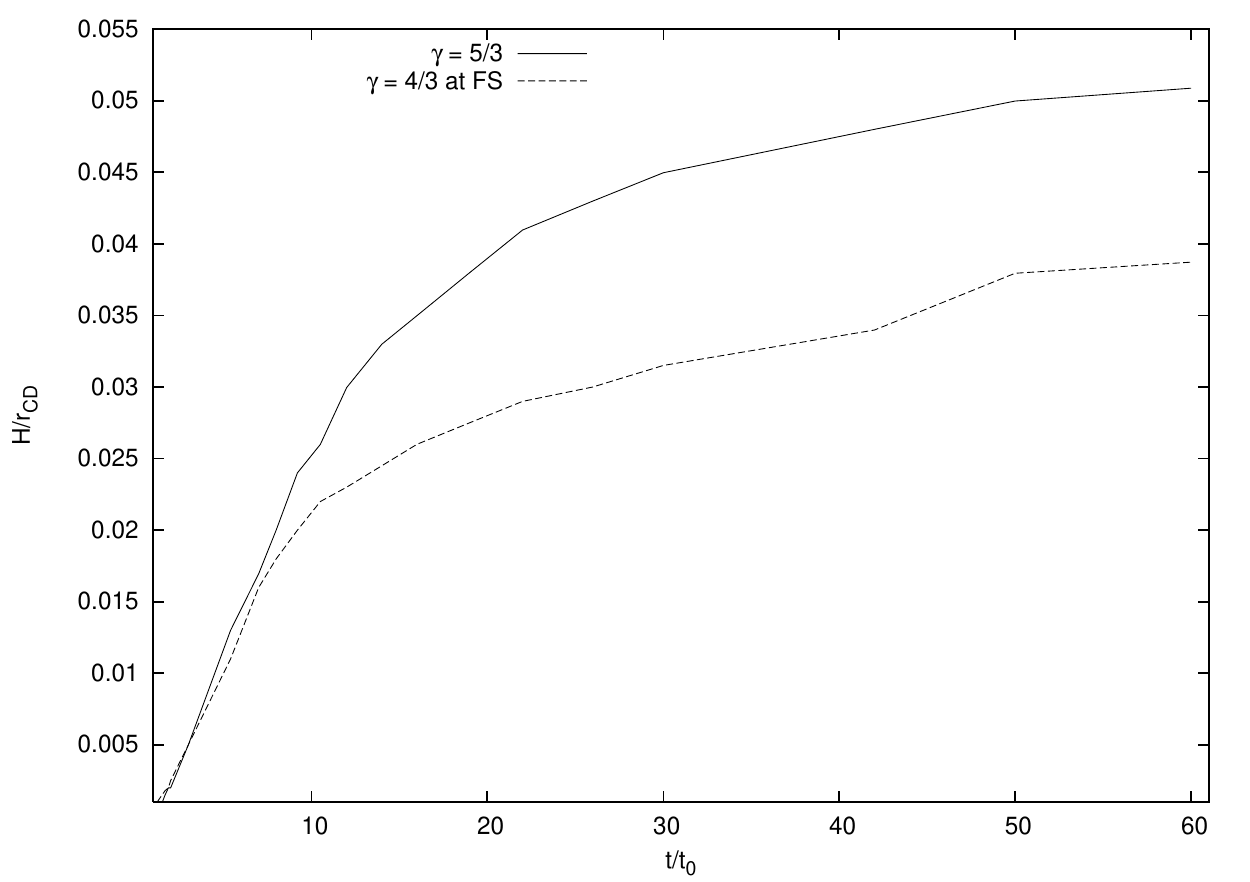}
\caption{The protrusion of RT structures measured in our simulations in units of the corresponding self-similar position of the contact discontinuity, whose surface is directly modified by the instabilities; the uniform $\gamma = 5/3$ and the hybrid cases are compared ($f(r) \sim 0.1$). The cosmic rays affect the elongation of the instabilities from the very beginning by a factor of the order of $20\%$.}
\label{layer_comp_Rcd}
\end{figure}

For $s = 0$, a convenient case to consider when analyzing the growth of RT structures is $n = 7$, for which the width of the interaction region is large, thus the elongation of the instabilities can be studied for a longer time
and the forward shock remains unaffected. 
The effect of changing the equation of state on the growth of the instabilities is represented in Fig.~\ref{fp_comp2}. 
It shows how far the RT fingers reach during their growth (100 yr) and when they are fully developed (500 yr). 
At 500 yr, the fraction of ejecta material transported well into the shocked ISM
(at $r/r_{\rm CD} = 1.05$, say) is lower in the hybrid case.
Nevertheless, the outermost ejecta reach nearly as far, and very close to the forward shock.

In the hybrid case, the position of the reverse shock is unaffected by a relativistic gas at the forward shock,
and the departure from the self-similar position is less than $2\%$. 
If the rela\-tivistic gas is uniformly distributed across the whole interaction region,
the departure of the reverse shock from the self-similar position is of the order of $6\%$.
Similar deviations have been found by \citet{be01}, who considered however 
a relatively small angular region in the cases $(n,s) = (7,2)$ or $(n,s) = (12,0)$. 
The bumps in $f(r)$ found at $r/r_{\rm CD} \sim 1.05$ originate in the RT structures themselves,
independent of the $\gamma$ distribution.
In the two cases $\gamma = 5/3$ and $\gamma = 4/3$, the width of the shocked ISM region ($r_{\rm FS}/r_{\rm CD}  - 1$) shrinks from 0.18 to 0.10. Comparable shrinking has been found in the angle-average analysis of $X$-ray observations of Tycho's SNR \citep{w05}.

As shown in Fig.~\ref{layer_comp_Rcd}, the elongation of the RT structures in the hybrid case
decelerates relative to the non-relativistic case. The deceleration is caused by the higher density gradient ahead of the contact discontinuity (cf. Fig.~\ref{d_comp3} and Fig.~\ref{ini}). 
This can be easily understood because the RT growth rate $\sigma$ depends through $A$ on the average ratio of the two densities
in the interaction region, i.e., $\rho_{\rm ej} / \rho_{\rm ISM}$ (cf. Fig.~\ref{d_comp3}),
thus lower density ratios produce lower growth rates (see Fig.~\ref{layer_comp_Rcd}).
The higher compression in the ISM shocked region in the hybrid case produces the slowing down
shown in Figs.~\ref{fp_comp2} and \ref{layer_comp_Rcd}.
When $\gamma = 4/3$ on both sides as in \citet{be01}, $\rho_{\rm ej} / \rho_{\rm ISM}$ remains larger than 1 
and the RT fingers reach further than in the hybrid case at $f(r) \leq 0.05$.

The region which drives the instability at very early time is that part of the density profile where density decreases outwards (see Fig.~\ref{ini}). This is entirely on the ejecta side and therefore unaffected by changing $\gamma$ in the shocked ISM (see Fig.~\ref{fp_comp2}).
In other words, at a very early stage, when the instabilities grow exponentially, 
the process occurs locally in the outer ejecta and the cosmic rays cannot affect it. 
However, particle acceleration will be able to affect RT structures as soon as they reach the shocked ISM, 
since the density gradient of the shocked ISM is higher for higher compression ratios. 
For the reasons explained above, a phenomenological equation of the type of Eq.~\ref{RTgrowth}
can only qualitatively reproduce the global behaviour. We could not find a more accurate description in this case.
A comparison of Fig.~\ref{layer_comp_Rcd} with Fig.~7 of \citet{be01} might indicate that 
the large fluctuations they observed in the late-time radial extent of RT structures  
are not an intrinsic property of the instability but a result of the too small physical size of the region considered.

\section{Conclusions}\label{conc}

We have adapted the AMR code RAMSES, which is based on a second order Godunov method in an expanding frame called ``supercomoving coordinates'', to follow the evolution of SNRs. In this approach, not only the space-time variables have been modified but also the hydrodynamics variables. The comoving coordinate system allows an eighth of the total volume of the SNR to be described, which is much larger than previously considered. A longer time interval can be investigated, of the order of thousands of years, until the transition to the asymptotic Sedov-Taylor phase. Such a large volume allows the convective instabilities to be modeled more accurately since it considers not only the shortest wavelengths, which have the greatest growth rate, but also the longer wavelengths, which grow more slowly but make a significant contribution to the morphology of the SNR over time-scales of thousands of years. Our larger spatial sampling allows a statistically more accurate description of the instabilities.

The great advantage of the method adopted here is that it minimizes the velocity of the fluid relative to the grid, and the numerical diffusion at the contact discontinuity, where the instability should develop. In the comoving reference frame, the contact discontinuity, in the absence of distortions due to convective instabilities and before the transition to the Sedov-Taylor phase, would be exactly stationary in time. The analytical non-inertial term resulting from this coordinate transformation is strictly equivalent to a gravitational acceleration: the Euler equations have therefore been integrated with the corresponding additional source terms.
 
The elongation of the Rayleigh-Taylor structures slowly reaches the asymptotic behaviour $t^\lambda$, by direct solution of the self-similar theory according to the assumption of a self-similar acceleration instead of a constant acceleration.

We have presented a simple way to numerically investigate the effect of efficient particle acceleration at the forward shock, approximating the shocked ISM as a relativistic gas and the shocked ejecta as a non-relativistic gas, with a different adiabatic exponent in each fluid.
The density behind the forward shock may be higher than at the reverse shock in that case. A deceleration in the protruding of RT structures is caused by the higher compression of the shocked ISM. The conclusion of \citet{be01} that the RT fingers can travel very close to the shock in the presence of accelerated particles remains valid. The elongation of the instabilities has here been more precisely quantified. We propose that this will allow us to understand why ejecta are found very close to the blast wave in the remnant of SN 1006 \citep{ch08}. The set-up of the code presented here will allow future studies of the back-reaction of particle acceleration on the SNR evolution. 

\begin{acknowledgements}
The numerical simulations have been performed with the DAPHPC cluster at CEA, DSM, for which the technical assistance is gratefully acknowledged. FF wishes to thank for useful discussions K. Kifonidis, E. Muller, A. Poludnenko. The work of FF was supported by CNES (the French Space Agency) and was carried out at Service d'Astrophysique, CEA/Saclay and partially at the Center for Particle Astrophysics and Cosmology (APC) in Paris. The authors are also grateful to the anonymous referee for the useful suggestions. 

\end{acknowledgements}


\end{document}